\PassOptionsToPackage{unicode}{hyperref}
\PassOptionsToPackage{hyphens}{url}
\PassOptionsToPackage{dvipsnames,svgnames,x11names}{xcolor}

\documentclass[
  12pt]{article}

\usepackage{amsmath,amssymb}
\usepackage{iftex}
\ifPDFTeX
  \usepackage[T1]{fontenc}
  \usepackage[utf8]{inputenc}
  \usepackage{textcomp}
\else 
  \usepackage{unicode-math}
  \defaultfontfeatures{Scale=MatchLowercase}
  \defaultfontfeatures[\rmfamily]{Ligatures=TeX,Scale=1}
\fi
\usepackage{lmodern}
\IfFileExists{upquote.sty}{\usepackage{upquote}}{}
\IfFileExists{microtype.sty}{
  \usepackage[]{microtype}
  \UseMicrotypeSet[protrusion]{basicmath}
}{}
\makeatletter
\@ifundefined{KOMAClassName}{
  \IfFileExists{parskip.sty}{
    \usepackage{parskip}
  }{
    \setlength{\parindent}{0pt}
    \setlength{\parskip}{6pt plus 2pt minus 1pt}}
}{
  \KOMAoptions{parskip=half}}
\makeatother
\usepackage{xcolor}
\usepackage{longtable,booktabs,array}
\usepackage{calc}
\usepackage{etoolbox}
\usepackage{graphicx}
\makeatletter
\def\maxwidth{\ifdim\Gin@nat@width>\linewidth\linewidth\else\Gin@nat@width\fi}
\def\maxheight{\ifdim\Gin@nat@height>\textheight\textheight\else\Gin@nat@height\fi}
\makeatother
\setkeys{Gin}{width=\maxwidth,height=\maxheight,keepaspectratio}
\makeatletter
\def\fps@figure{htbp}
\makeatother

\usepackage{caption}
\usepackage{subcaption}
\usepackage{float} 

\ifLuaTeX
  \usepackage{selnolig}  
\fi
\usepackage[]{natbib}
\usepackage{bookmark}

\IfFileExists{xurl.sty}{\usepackage{xurl}}{} 
\usepackage{amsthm}
\usepackage{mathtools}
\usepackage{mathrsfs}

\DeclareMathOperator{\dist}{dist}
\DeclareMathOperator{\median}{median}
\newcommand{\W}{\mathbf{W}}

\newcommand{\anon}{1}
\newcommand{\blind}[2]{\if1\anon #1\else #2\fi}

\hypersetup{
  pdftitle={Modeling Disease Transmission Dynamics Using Wastewater Data and Population-Weighted Distance Methods},
  pdfkeywords={Respiratory syncytial virus (RSV); Bayesian renewal model; Effective reproduction number; Population Extended Hausdorff Distance; State-space model; Spatial coupling},
  colorlinks=true,
  linkcolor={blue},
  filecolor={Maroon},
  citecolor={Blue},
  urlcolor={Blue},
  pdfcreator={LaTeX}}
\if1\anon
  \hypersetup{pdfauthor={Jose R. Palacio; Katherine B. Ensor; Julia Schedler; Rebecca Schneider; Kaavya Domakonda; Loren Hopkins; Lauren B. Stadler; Sharmila Bhandari}}
\fi

\begin{document}

\def\spacingset#1{\renewcommand{\baselinestretch}
{#1}\small\normalsize} \spacingset{1}

\if1\anon
{
  \title{\bf Modeling Disease Transmission Dynamics Using Wastewater Data and Population-Weighted Distance}
  \author{Jose R. Palacio$^{1}$\thanks{Corresponding author: \texttt{jrp16@rice.edu}},
    Katherine B. Ensor$^{1}$,
    Julia Schedler$^{2}$,\\
    Rebecca Schneider$^{3}$,
    Kaavya Domakonda$^{3}$,
    Loren Hopkins$^{1,3}$,\\
    Lauren B. Stadler$^{4}$,
    and
    Sharmila Bhandari$^{3}$\\[.3cm]
    {\small $^{1}$Department of Statistics, Rice University, 6100 Main St, Houston, TX, USA}\\
    {\small $^{2}$Statistics Department, California Polytechnic State University, San Luis Obispo, CA}\\
    {\small $^{3}$Houston Health Department, Houston, TX, USA}\\
    {\small $^{4}$Department of Civil and Environmental Engineering, Rice University, Houston, TX, USA}}
  \maketitle
} \fi

\if0\anon
{
  \bigskip
  \bigskip
  \bigskip
  \begin{center}
    {\LARGE\bf Modeling Disease Transmission Dynamics Using Wastewater Data and Population-Weighted Distance Methods}
\end{center}
  \medskip
} \fi

\bigskip
\begin{abstract}
Wastewater-based epidemiology (WBE) is an effective, noninvasive tool for tracking community-level circulation of respiratory viruses, yet standard renewal models treat each wastewater treatment plant (WWTP) in isolation, even though respiratory syncytial virus (RSV) spreads through contact networks that cross service-area boundaries. Using weekly data from $m = 15$ Houston WWTPs, we extend a single-plant Bayesian renewal model by coupling neighboring WWTPs via a spatial weight matrix built using a novel Population Extended Hausdorff Distance, defined as a directional distance between WWTPs. This distance measures how far a WWTP must reach to access a meaningful share of a neighboring WWTP's residents. A single parameter $\rho$ controls spatial mixing in the plant-level growth rate, allowing us to recover the effective reproduction number $R_{it}$. We find that $\rho$ is estimated well above zero, indicating substantial cross-WWTP transmission that the independent-plant model cannot capture. Spatial coupling reshapes plant-level $R_{it}$ relative to the independent-plant baseline ($\rho = 0$) while leaving inferred infection trajectories essentially unchanged.
\end{abstract}

\noindent%
\vfill

\newpage
\spacingset{1.8} 

\section{Introduction}\label{sec:introduction}

Wastewater-based epidemiology (WBE) has established itself as one of the pillars of circulating pathogen surveillance in urban populations. Its noninvasive, cost-effective nature, and its independence from health care-seeking behaviors, as it simultaneously captures symptomatic, asymptomatic, and underdiagnosed cases, make it an ideal complement to clinical encounter records for understanding and estimating the prevalence of a virus in specific population segments. Studies such as those by \citet{stadler2020wastewater, hopkins2023citywide, ensor2025nonlinear} and \citet{palacio2026inferring} demonstrate how it is possible to track increases in cases within a population and estimate, in relative terms, the number of new infections during specific time periods, with trajectories that closely mirror those of reported clinical encounters \citep{ensor2026understanding}.

However, viral load measurements alone cannot identify the absolute number of infections, because the observed concentrations depend on WWTP-specific factors such as excretion rates, dilution, flow rates, and losses throughout the sewer network. For this reason, the WBE literature has converged on the effective reproduction number, $R_t$, as the primary metric for transmission \citep{huisman2022estimation, champredon2024ern}. The effective reproduction number is a relative and dimensionless quantity, $R_t$ is invariant to these scaling factors and allows for a direct epidemiological interpretation ($R_t > 1$ indicates growth, while $R_t < 1$ indicates decline). In the multi-WWTP context addressed in this study, this property is also indispensable: absolute viral loads across Houston's treatment plants are not directly comparable with one another, whereas $R_{it}$ is; this enables spatial coupling between WWTP areas that yields a coherent epidemiological interpretation. The broader case for wastewater surveillance as a public health instrument has been made in recent perspective and review work \citep{diamond2022wastewater}, and the demonstrated capacity of WBE to capture subpopulation dynamics (for example, school-specific outbreaks within a citywide network) underscores the value of multiresolution monitoring \citep{wolken2023wastewater}.

Nevertheless, studies such as that by \citet{palacio2026inferring} assume that each plant operates in isolation, modeling each WWTP service area as an independent epidemic. This constitutes a highly limiting assumption, as RSV transmission operates through contact networks (schools, workplaces, public transportation, mixed households). These networks do not align with the geometric boundaries of WWTP service areas, boundaries defined by sanitary engineering rather than epidemiology. For instance, a primary case in zone $A$ may generate a secondary case in zone $B$ if individuals cross these boundaries during their infectious window. Ignoring such potential scenarios precludes leveraging shared information among neighboring WWTPs, resulting in $R_{it}$ trajectories that appear more weakly coupled than the actual epidemiological dynamics would suggest. In this work, we extend the Bayesian renewal model to epidemiologically couple Houston's various wastewater treatment plants while preserving the interpretability of the single-plant model proposed by \citet{palacio2026inferring}.

Modeling infectious diseases within a spatial framework is not uncharted territory in the literature. Various studies demonstrate the coupling of neighboring areas through a component that captures how transmission flows across administrative boundaries. For example, the endemic--epidemic framework introduced by \citet{held2005statistical} decomposes the observed incidence in each spatial unit into a additive combination of a unit-specific autoregressive component and a neighbor-specific epidemic component. \citet{meyer2014powerlaw} extend this construction \citep[building on][]{paul2008multivariate, heldpaul2012modeling} by replacing binary adjacency with weights that decay according to a power law based on neighborhood order, motivated by empirical evidence indicating that short-range human mobility follows a power law. In parallel, within the field of Bayesian disease mapping, modern formulations, such as the BYM2 model by \citet{riebler2016intuitive} have demonstrated how to modulate the intensity of spatial coupling using a single scalar parameter that convexly weights a spatial structure against an independent alternative. We can, also, find works that incorporated space-time extensions through a scalar coupling parameter determining the transmission between neighboring areas such as the model ZS-CMSNB in \citep{douwesschultz2022zerostate} applied on dengue in Rio de Janeiro. However, all these methods define relationships between neighborhoods by the layout of regions on the map, either via binary adjacency between regions or via graph-theoretic distances derived from it, ignoring how the population is distributed within each region. This work addresses this limitation by introducing a population-weighted directional distance within a spatially coupled Bayesian renewal framework.

We now outline the structure of the manuscript. In Section \ref{sec:methods}, we describe the data; the construction of the Population Extended Hausdorff Distance (PEHD) matrix and the spatial weight matrix $\W$; the generation and shedding kernels; the spatially coupled latent process for the growth rate $\lambda_{it}$; and the spatial renewal equation; the viral load observation model; the priors and posterior inference procedure; and the two-stage strategy. Next, in Section \ref{sec:results}, we present convergence diagnostics, the posterior distribution of the spatial coupling parameter $\rho$, plant-specific scale estimates, and the inferred trajectories of $R_{it}$ and $I_{it}$ under both input variants (Filter and Data), comparing them against the spatially independent baseline at $\rho = 0$. In Section \ref{sec:discussion}, we interpret the principal findings, including the magnitude of coupling between WWTPs, the robustness of the inferred incidence to $\rho$, the implications for spatial wastewater surveillance, and the methodological limitations of the framework.

\section{Statistical Framework and Methodology}\label{sec:methods}

Our analysis is based on a Bayesian renewal model for multiple WWTPs. Building on the single WWTP renewal framework of \citet{palacio2026inferring}, we extend it to $m = 15$ WWTPs by introducing a spatial weight matrix $\W$ that couples neighboring WWTPs through the latent growth rate. A single mixing parameter $\rho \in [0, 1]$ controls the strength of coupling between WWTPs by weighting each WWTP's own past against a population weighted average of its neighboring WWTPs. We perform joint posterior inference for $\rho$ and the WWTP specific scale parameters using Hamiltonian Monte Carlo in Stan, while signal extraction from the viral load trajectories at the WWTP level is carried out independently with the \texttt{MARSS} R package \citep{holmes2012marss}.

\subsection{Data and study area}\label{sec:data}

We analyze weekly RSV viral load expressed in billions of genome copies per day (B gc/day) at $m = 15$ wastewater sources operated by the Houston Health Department: five individual treatment plants and ten physically pooled samples, each combining two to five contributing plants. For brevity, we refer to all $m$ sources as  "plants" throughout the manuscript. For each plant $i \in \{1,\dots, m\}$ and week $t \in \{1,\dots, T\}$, observed concentrations $C_{it}$ (gc/L) are converted to plant-specific daily loads using the median daily influent flow $\bar F_i$ (L/day) of the contributing plants:

\begin{equation}\label{eq:load-conv}
y_{it}  =  C_{it}\bar F_i \times 10^{-9},
\qquad i =  1,\dots, m,\ \ t = 1,\dots, T.
\end{equation}

The $m$ plants jointly serve $\sum_i N_i \approx 2.16 \times 10^6$ residents, essentially the entire wastewater-served population of Houston, with individual WWTPs spanning nearly two orders of magnitude, from $\sim 1.1\times 10^4$ to $\sim 5.8\times 10^5$ residents. Twenty-four-hour composite samples were collected weekly from January 2023 to May 2025; weeks without sampling are treated as missing.

Each pooled sample is constructed by combining equal volumes from two to five contributing plants' replicate samples. Pool membership is fixed throughout the study period, and each individual plant contributes to at most one pool. The service area polygon $\Omega_i$ of a pooled source is taken as the geometric union of the contributing plants' service area polygons, and the corresponding population as the sum of their populations. The full pooling protocol is described in \citet{ensor2025nonlinear}.

Plant-level served populations $N_i$ are estimated from a spatial overlay of each service area polygon $\Omega_i$ with the 2022 U.S. Census tract boundaries. The wastewater service area shapefile is obtained from Houston Public Works and intersected with the Census tract polygons in ArcGIS Pro to produce, for each tract $c$ and plant $i$, an intersection feature whose area-share fraction $\alpha_{c, i} \in [0,1]$ is the proportion of the tract's area that falls inside $\Omega_i$. Tract populations $P_c$ are taken from the 2024 American Community Survey five-year estimates and joined to the intersection features by Census GEOID. Intersections with $\alpha_{c, i} < 0.05$ are discarded to suppress boundary-overlap artifacts. The plant-level population is then the area-share-weighted aggregate over the retained tracts:

\begin{equation}\label{eq:plant-pop}
N_i  =  \sum_{c} \alpha_{c, i} P_c.
\end{equation}
These populations $\{N_i\}_{i=1}^m$ enter the methodology in two places: (i) anchoring the initial incidence $I_{i1}$ via the citywide weekly RSV case count (Section \ref{sec:incidence}), and (ii) constructing the spatial weight matrix $\W$ (Section \ref{sec:pehd}).

\begin{figure}[t]
\centering
\includegraphics[width=0.85\linewidth]{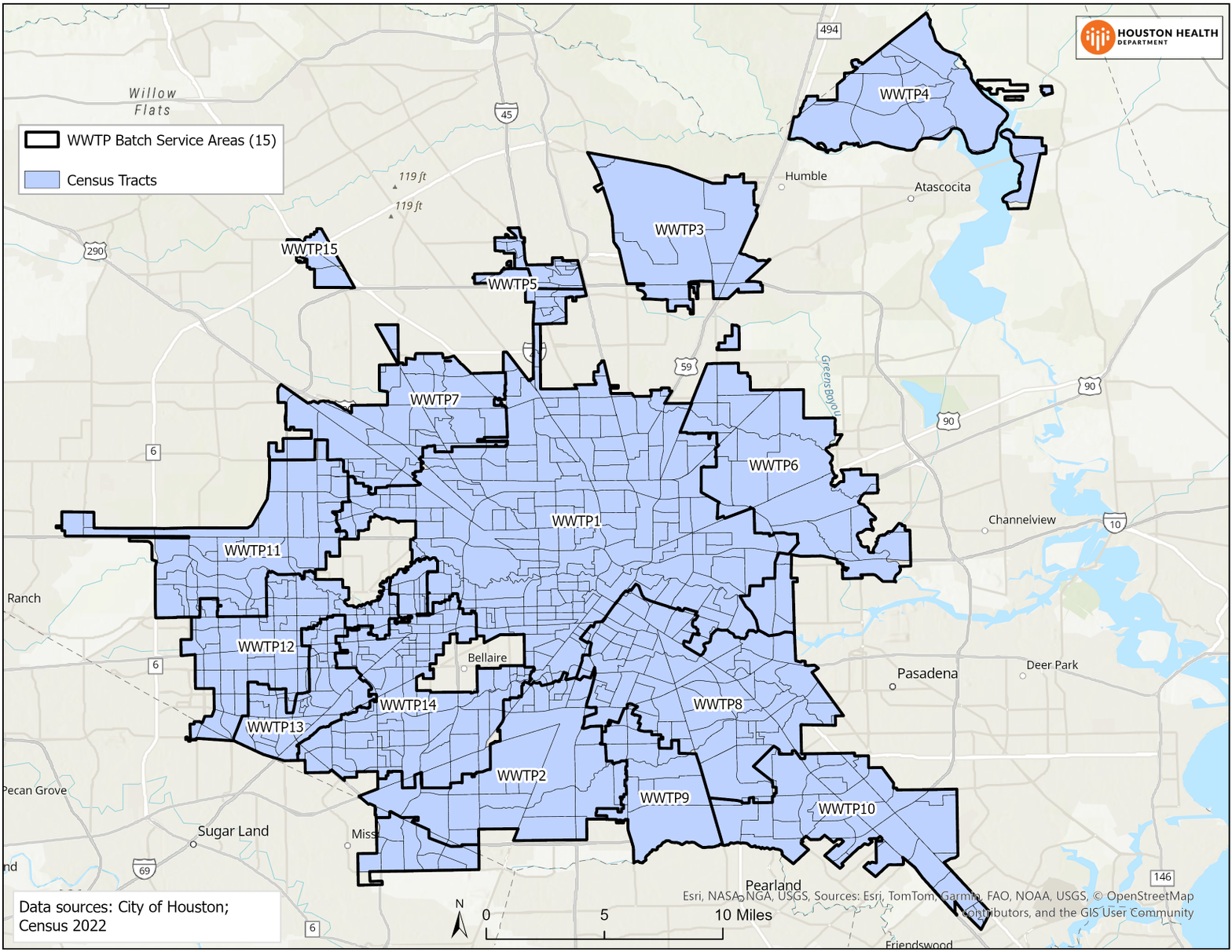}
\caption{Service area polygons of the $m = 15$ Houston WWTPs used in this study (heavy outlines), overlaid on the underlying census tract geometry (light fill). Plant-level populations $N_i$ are obtained by allocating tract populations to each WWTP via the area-share fractions $\alpha_{c, i}$ of equation \eqref{eq:plant-pop}. Sources: City of Houston wastewater service area shapefiles; U.S.\ Census Bureau 2022 tract boundaries.}
\label{fig:houston-map}
\end{figure}

\subsection{Population Extended Hausdorff Distance and spatial weight matrix \texorpdfstring{$\W$}{W}}\label{sec:pehd}

The spatial extension of the renewal equation (Section \ref{sec:modelspec}) requires an $m \times m$ spatial weight matrix $\W$ encoding whether and how each plant's latent infections inform its neighbors. Spatial coupling is essential when the relevant contact network spans multiple WWTPs: people who live in WWTP $i$ but spend significant time in WWTP $j$ contribute contacts to both, and the infections those contacts produce should propagate accordingly. We construct $\W$ in three steps: a population-derived directional distance we call the Population Extended Hausdorff Distance (PEHD) matrix (Section \ref{sec:pehd-def}), a Gaussian (radial basis function) kernel transformation (Equation \eqref{eq:rbf}), and row normalization (Equation \eqref{eq:row-norm}).

\subsubsection{Population Extended Hausdorff Distance Matrix}\label{sec:pehd-def}

The natural starting point for measuring distance between two spatial regions is the classical (one-sided) Hausdorff distance from compact set $A$ to compact set $B \subset \mathbb{R}^2$,
$$
d_H(A \to B) = \sup_{a \in A} \inf_{b \in B} \|a - b\|_2,
$$
i.e.\ the worst-case point-to-set distance. This is a purely geometric measure: it ignores how the population of $B$ is distributed inside the polygon. In an epidemiological mixing matrix, however, what matters is not  "how far is the farthest point of $B$ from $A$?" but rather  "how far does $A$ have to reach to access a meaningful share of the people living in $B$?" A WWTP whose population is concentrated near the shared boundary with a neighbor is, for transmission purposes, much closer to that neighbor than its geometric extent suggests. On the other hand, a WWTP whose population is concentrated far from the shared boundary is much farther than the polygon geometry would imply.

We therefore generalize the Hausdorff distance \citep{min2007extended, schedler2020advances} along two axes: replace the supremum with a quantile over the population substrate of $B$, and symmetrize over the two directions. Formally, let $\Omega_j \subset \mathbb{R}^2$ be plant $j$'s service area polygon (projected to UTM zone 15N, meters) and $\mathrm{d}P_j(\mathbf x)$ the population measure of $\Omega_j$ derived from the tract-level allocation $\{\alpha_{c, j}, P_c\}$ of equation \eqref{eq:plant-pop}. For an ordered pair of plants $(i, j)$, the population median effective distance is the smallest distance from the boundary of $\Omega_i$ that captures at least half of plant $j$'s population:

\begin{equation}\label{eq:asym-pehd}
\tilde d_{ij}
 =
\inf\left\{r \geq 0  :
\frac{\int_{\Omega_j} \mathbf 1\{\dist(\mathbf x,\Omega_i)\leq r\}\mathrm{d}P_j(\mathbf x)}
     {\int_{\Omega_j} \mathrm{d}P_j(\mathbf x)}
 \geq  \tfrac{1}{2}\right\},
\end{equation}
where the integration variable $\mathbf x \in \Omega_j$ ranges over the target polygon, and $\dist(\mathbf x, \Omega_i) = \inf_{\mathbf y \in \Omega_i}\|\mathbf x - \mathbf y\|_2$ is the Euclidean point-to-set distance from $\mathbf x$ to the source polygon $\Omega_i$ (zero if $\mathbf x \in \Omega_i$).

We refer to $\tilde d_{ij}$ as a \emph{directional distance} between WWTPs: it is, in general, asymmetric in the ordered pair $(i, j)$ because the population substrates $\mathrm{d}P_i$ and $\mathrm{d}P_j$ differ. This asymmetric form is the core quantity of the construction; the symmetrization in equation \eqref{eq:sym-pehd} below aggregates the two directions into a single per-pair weight used by the downstream kernel. Collecting these directional distances yields into a asymmetric matrix $\widetilde{\mathbf D} = \left(\tilde d_{ij}\right)^{m}_{i,j=1}$.

\begin{figure}[H]
\centering
\includegraphics[width=\linewidth]{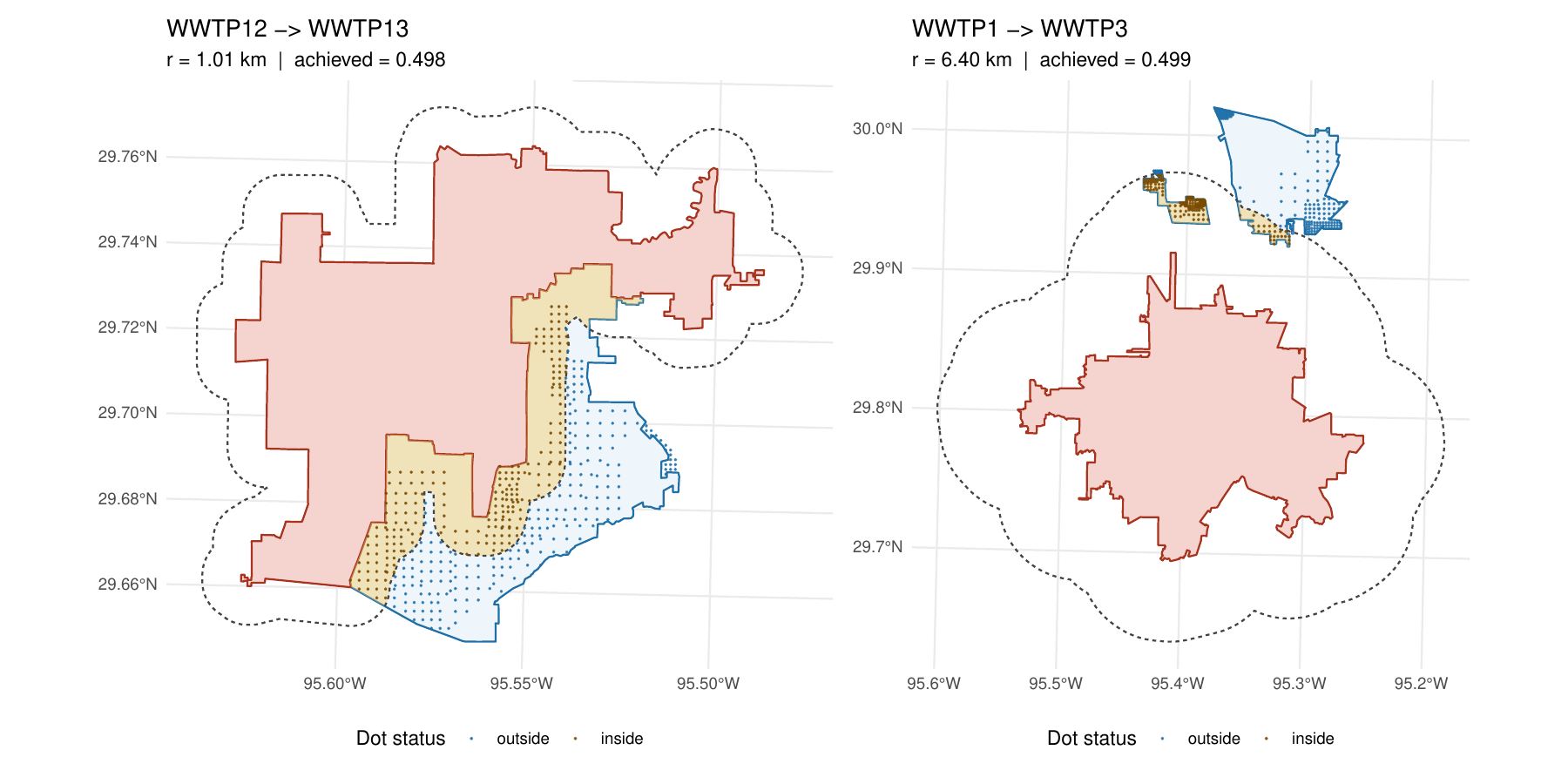}
\caption{Two examples of the asymmetric population median effective distance $\tilde d_{i \to j}$ from equation \eqref{eq:asym-pehd}: a close pair (WWTP12 $\to$ WWTP13, left) and a distant pair (WWTP1 $\to$ WWTP3, right). Red: sewershed $\Omega_i$. Blue: sewershed $\Omega_j$, with residents as dots (gold inside, blue outside the dashed contour). The dashed contour is the expanding distance threshold around $\Omega_i$'s boundary and it grows outward until half of $j$'s residents are captured.}
\label{fig:pehd-cross-pairs}
\end{figure}

The construction generalizes to any population quantile $q \in (0,1)$ by replacing $1/2$ in \eqref{eq:asym-pehd} with $q$; we write $\tilde d_{ij}^{(q)}$ for the resulting family. The choice $q = 1/2$ is the natural default: it is robust to outlying residents at either polygon's edges, and corresponds to the distance below which a typical resident of $\Omega_j$ can reach $\Omega_i$. We use $q = 1/2$ throughout and drop the superscript.

Because the population substrate of $\Omega_j$ differs from $\Omega_i$, $\widetilde{\mathbf D}_{ij}$ is in general asymmetric. We symmetrize by the elementwise maximum,
\begin{equation}\label{eq:sym-pehd}
\mathbf D = (d_{ij})_{i,j=1}^{m}, \qquad
d_{ij} = \max(\tilde d_{ij}, \tilde d_{ji}),
\end{equation}
and refer to $\mathbf D$ as the \emph{Population Extended Hausdorff Distance} (PEHD) matrix. By construction, $\mathbf D$ satisfies the three properties required by the downstream construction: (i) symmetry, $d_{ij} = d_{ji}$; (ii) nonnegativity, $d_{ij} \geq 0$; and (iii) zero diagonal, $d_{ii} = 0$, since plant $i$ trivially captures its own population median at $r = 0$. These three properties are all that the Gaussian RBF transformation \eqref{eq:rbf} and the row normalization \eqref{eq:row-norm} require of $d_{ij}$; no further structural assumption is used in any step that follows.

In implementation, the population measure $\mathrm{d}P_j$ is approximated by a Monte Carlo dot mesh: $M_j = \lfloor N_j / \nu \rfloor$ points are placed uniformly inside $\Omega_j$ proportional to tract-level populations, with $\nu = 200$ people per dot. The integral in \eqref{eq:asym-pehd} then becomes a finite count of dots whose polygon distance to $\Omega_i$ is at most $r$, and the infimum reduces to the $\lceil M_j / 2 \rceil$th order statistic of those distances. With $m = 15$ plants, the full PEHD matrix is computed once as a preprocessing step.

\subsubsection{From distances to spatial weights}\label{sec:weights}

The PEHD matrix is transformed into the spatial weight matrix $\W$ in two steps. First, distances are mapped to nonnegative similarities via the Gaussian radial basis function (RBF) kernel,
\begin{equation}\label{eq:rbf}
K_{ij}  =  \exp\left(-\frac{d_{ij}^{2}}{2\sigma_K^{2}}\right),
\qquad \sigma_K^{2}  =  \tfrac{1}{2}\median\left\{d_{ij}^{2}: i \neq j\right\},
\end{equation}
with bandwidth set by the median heuristic on squared distances \citep{gretton2012kernel}. The Gaussian RBF is monotonically decreasing in $d_{ij}$, infinitely differentiable, and bounded in $(0,1]$, so that closer WWTP pairs receive larger similarity values, the diagonal $K_{ii}$ equals one by construction, and the resulting kernel matrix is smooth in any function of $\W$. The median heuristic, in turn, is scale-invariant and robust to outlying plant pairs; \citet[Theorem 1]{garreau2017large} establish its asymptotic consistency.

Second, we zero the diagonal: each plant's own contribution to its dynamics enters through the $(1-\rho)$ term in Section \ref{sec:modelspec}, so the spatial component must carry only the contributions of neighbors, and we divide each row by its sum:
\begin{equation}\label{eq:row-norm}
K_{ii} \leftarrow 0, \qquad
\W_{ij}  =  \frac{K_{ij}}{\sum_{k=1}^m K_{ik}}.
\end{equation}
After this step, each row of $\W$ sums to one: $\sum_j \W_{ij} = 1$ for every plant $i$. The entry $\W_{ij}$ has a clean interpretation as the relative weight that plant $i$'s spatial mean assigns to plant $j$. By normalizing the rows, we preserve the natural scale of $\lambda$: the spatial mean in \eqref{eq:lambda-mix} stays a weighted average of neighbors' growth rates on the same scale as the plant's own growth rate, so the convex combination with weight $\rho$ remains directly interpretable.

Row normalization breaks the symmetry inherited from $\mathbf{K}$: although $K_{ij} = K_{ji}$ by construction, the row sums $\sum_k K_{ik}$ differ across plants in dense versus sparse neighborhoods, so $\W_{ij} \neq \W_{ji}$ in general. This is intentional: $\W_{ij}$ encodes the share of plant $i$'s spatial mean assigned to plant $j$, a directional quantity that depends on plant $i$'s overall neighborhood structure.

\begin{figure}[H]
\centering
\includegraphics[width=0.8\linewidth]{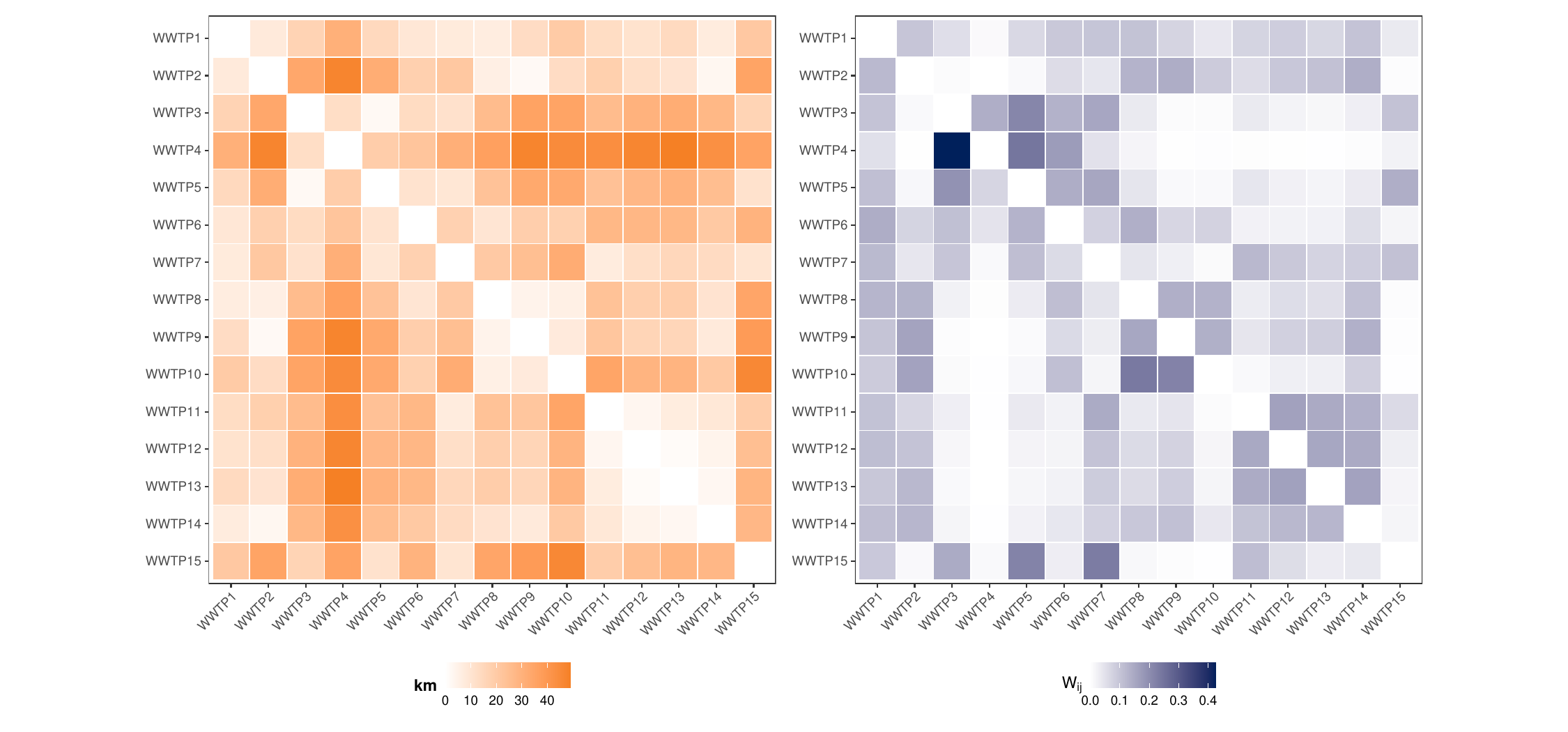}
\caption{Spatial structure of the WWTP network in Houston. Left: Population Extended Hausdorff Distance matrix $\mathbf D$ for the 15 plant WWTPs, in km (equation \eqref{eq:sym-pehd}); off-diagonal entries range from approximately $2$ km between geographically adjacent WWTPs to $\sim 49$ km between the most distant pairs. Right: resulting spatial weight matrix $\W$, after the Gaussian RBF kernel transformation of equation \eqref{eq:rbf} and the row normalization of equation \eqref{eq:row-norm}. In both heatmaps, the diagonal vanishes by construction; plants are ordered WWTP1 through WWTP15.}
\label{fig:matrices}
\end{figure}

\subsection{Generation and shedding kernels}\label{sec:kernels}

Transmission and observation delays are represented by two kernels: the generation interval $\{g_\tau\}_{\tau=1}^{G}$ and the shedding profile $\{s_\tau\}_{\tau=0}^{S}$. Both are modeled as continuous-time Gamma distributions and discretized into weekly lags via cumulative distribution function (CDF) differences.

The generation interval governs the renewal of infections, the distribution of time elapsed between a primary case's infection and the secondary cases it produces. We follow \citet{vink2014serial}, who reviewed serial interval estimates for several respiratory infections and derived a pooled RSV mean of 7.5 days with SD 2.1 days, and, converted to the weekly time step of the model, parameterize the generation interval with $\mu_G \approx 1.07$ weeks and $\sigma_G \approx 0.30$ weeks. Letting $F_\Gamma(\cdot;\kappa_G,\theta_G)$ denote the Gamma CDF with shape $\kappa_G = \mu_G^2/\sigma_G^2$ and scale $\theta_G = \sigma_G^2/\mu_G$,
\begin{equation}\label{eq:gen-kernel}
g_\tau  \propto
F_\Gamma\left(\tau + \tfrac12;\kappa_G,\theta_G\right)
- F_\Gamma\left(\tau - \tfrac12;\kappa_G,\theta_G\right),
\quad \tau = 1,\dots, G,
\end{equation}
with support truncated at $G$ weeks and weights re-normalized to sum to one. Lag zero is excluded to reflect the biological delay between primary and secondary infections \citep{fraser2007estimating}.

The shedding profile maps each infection to its viral contribution to wastewater over time. \citet{okiro_duration_2010} synthesize adult and pediatric RSV viral shedding studies into a pooled mean of 4.5 days (SD 2.0 days). Converted to weeks, we adopt these values to set $\mu_S \approx 0.66$ weeks and $\sigma_S \approx 0.29$ weeks. Unlike the generation interval, the shedding profile retains lag zero, because RSV shedding can begin within the first few days of illness \citep{hall2001respiratory, devincenzo2005respiratory}:

\begin{equation}\label{eq:shed-kernel}
s_\tau  \propto
F_\Gamma\left(\tau + \tfrac12;\kappa_S,\theta_S\right)
- F_\Gamma\left(\max\{\tau - \tfrac12,0\};\kappa_S,\theta_S\right),
\quad \tau = 0,\dots, S,
\end{equation}
with support truncated at $S$ weeks and weights re-normalized to sum to one. In both cases, we truncate at two weeks.

The two kernels differ in their spatial behavior. A contact between a primary case at plant $i$ and a secondary case at plant $j \neq i$ at lag $\tau$ weeks is governed by the same generation interval distribution as a within-plant contact, but the resulting infection contributes to plant $j$'s incidence rather than plant $i$'s. The generation kernel itself is therefore plant-invariant; cross-plant routing of secondary infections is handled separately by the spatial weight matrix $\W$ (Section \ref{sec:pehd}). The shedding profile, by contrast, is intrinsically plant-local: an infected individual's wastewater contribution flows to the plant whose sewer they are physically connected to, regardless of where the infection occurred, so no spatial mixing is required. 

\subsection{Model specification}\label{sec:modelspec}

The latent epidemiological process is modeled as a five-step chain. We first specify a spatially coupled Gaussian random walk on the unconstrained growth rate $\lambda_{it}$ (Section \ref{sec:lambda}); recover the effective reproduction number $R_{it}$ from $\lambda_{it}$ in closed form via the Euler--Lotka relation (Section \ref{sec:euler-lotka}); compute the expected weekly incidence $\iota_{it}$ through a spatially mixed renewal convolution (Section \ref{sec:renewal}); sample realized incidence $I_{it}$ from a continuous lognormal relaxation of the Poisson distribution (Section \ref{sec:incidence}); and finally map incidence to expected viral  load through a plant-specific shedding process and observe it under a  lognormal likelihood (Section \ref{sec:viral-load}). The single coupling parameter $\rho \in [0,1]$ enters both the random walk on $\lambda$ and the renewal convolution, so that across WWTP networks contact propagation manifests at both the latent rate and the incidence levels.

\subsubsection{Spatial random walk on the growth rate}\label{sec:lambda}

For each plant $i$, we model the unconstrained growth rate $\lambda_{it}$ as a Gaussian random walk with spatial mixing through $\W$. The initial state is a diffuse Gaussian centered at zero,
\begin{equation}\label{eq:lambda-init}
\lambda_{i1}  \sim  \mathcal{N}(0,\sigma_{\lambda}^{2}),
\end{equation}
and subsequent steps evolve as
\begin{equation}\label{eq:lambda-walk}
\lambda_{it} \mid \boldsymbol\lambda_{t-1}
 \sim
\mathcal{N}\left(\mu^{\lambda}_{it},\sigma_\epsilon^{2}\right),
\qquad t \geq 2,
\end{equation}
where the conditional mean is a convex combination of the plant's own past and a population-weighted average of its neighbors' pasts:
\begin{equation}\label{eq:lambda-mix}
\mu^{\lambda}_{it}
 =  (1-\rho)\lambda_{i, t-1}
 +  \rho \sum_{j=1}^{m} \W_{ij}\lambda_{j, t-1}.
\end{equation}
We fix $\sigma_{\lambda} = \sigma_\epsilon = 0.5$ throughout.

Equation \eqref{eq:lambda-mix} encodes the contact-driven hypothesis that the underlying drivers of $\lambda$ (population mixing, seasonality, behavior) are partially shared across neighboring WWTPs. The two limits are natural sanity checks. When $\rho = 0$, the spatial term vanishes and the process reduces to $m$ independent random walks, recovering the single-plant model of \citet{palacio2026inferring}. When $\rho = 1$, the plant's own past disappears and the conditional mean is purely the spatial average $\sum_j \W_{ij}\lambda_{j, t-1}$. Intermediate values interpolate between these two regimes, with $\rho$ controlling the relative weight of local versus regional dynamics.

\subsubsection{Recovering \texorpdfstring{$R_{it}$}{Rit}: the Euler--Lotka relation}\label{sec:euler-lotka}

The latent process operates on the unconstrained growth rate $\lambda_{it}$, but the quantity of epidemiological interest is the effective reproduction number $R_{it}$. The two are related by a closed-form Euler--Lotka root specific to a Gamma-distributed generation interval \citep{wallinga2007generation}. Let the generation interval be a continuous random variable $\tau \geq 0$ with density $g(\tau)$ of $\mathrm{Gamma}(\kappa, \theta)$, so that mean $\mu_G = \kappa\theta$ and variance $\sigma_G^2 = \kappa\theta^2$. For incidence growing as $i(t) \propto e^{\lambda t}$, the reproduction number $R$ associated with $\lambda$ satisfies
\begin{equation}\label{eq:euler-lotka}
R_{it}  =  \left(1 + \frac{\sigma_G^{2}}{\mu_G}\lambda_{it}\right)^{\mu_G^{2}/\sigma_G^{2}}.
\end{equation}

%

In the implementation, $\lambda_{it}$ is a per-week growth rate. Equation \eqref{eq:euler-lotka} is monotone increasing in $\lambda_{it}$, returns $R_{it} = 1$ at the equilibrium $\lambda_{it} = 0$, and is well defined for all $\lambda_{it} > -\mu_G / \sigma_G^{2}$, which covers the entire physically meaningful regime. We compute $R_{it}$ deterministically from $\lambda_{it}$ at every Hamiltonian Monte Carlo iteration; the posterior on $R_{it}$ is therefore induced by the posterior on $\lambda_{it}$.

\subsubsection{Spatially mixed renewal equation}\label{sec:renewal}

We define the spatially blended past infection trajectory at plant $i$ and week $t$ as a convex combination of plant $i$'s own infections and a population-weighted average of its neighbors':
\begin{equation}\label{eq:i-mix}
I^{\mathrm{mix}}_{it}
 =  (1-\rho) I_{it}  +  \rho \sum_{j=1}^{m} \W_{ij} I_{jt}.
\end{equation}
The same coupling parameter $\rho$ from Section \ref{sec:lambda} appears here: a primary infection in WWTP $j$ at week $t$ can generate a secondary infection in WWTP $i$ at week $t + \tau$ if a person from $j$'s WWTP contacted a person from $i$'s WWTP during their infectious window, and $\W_{ij}$ quantifies the intensity of that cross-WWTP contact. The expected weekly incidence is then the renewal convolution of $R_{it}$ with the blended trajectory through the generation kernel:
\begin{equation}\label{eq:renewal}
\iota_{it}
 =  R_{it} \sum_{\tau=1}^{G} g_\tau I^{\mathrm{mix}}_{i, t-\tau},
\qquad t \geq 2.
\end{equation}
When $\rho = 0$ the convolution acts only on the plant's own past and \eqref{eq:renewal} collapses to the standard single-plant renewal equation of \citet{fraser2007estimating}; when $\rho > 0$, information from neighboring WWTPs flows into the plant's expected incidence through the row $\W_{i,\cdot}$.

\subsubsection{Lognormal relaxation of the Poisson likelihood}\label{sec:incidence}

The natural distribution for weekly incidence counts $I_{it}$ given expected incidence $\iota_{it}$ is Poisson, but the discreteness of the Poisson is incompatible with the gradient-based Hamiltonian Monte Carlo sampling we use for posterior inference. We therefore relax the Poisson to a continuous lognormal whose parameters are chosen to match the Poisson's mean and variance, $\mathbb{E}[I_{it} \mid \iota_{it}] = \iota_{it}$ and $\mathrm{Var}[I_{it} \mid \iota_{it}] = \iota_{it}$:
\begin{equation}\label{eq:i-lognormal}
I_{it} \mid \iota_{it}
 \sim
\mathrm{Lognormal}\left(\log\iota_{it} - \tfrac{1}{2\iota_{it}},
\iota_{it}^{-1/2}\right),
\qquad t \geq 2.
\end{equation}
The resulting distribution preserves the Poisson's mean--variance equality to leading order in $\iota_{it}^{-1}$ while being smooth in $\iota_{it}$ and therefore amenable to gradient-based sampling.

The initial week requires an exogenous anchor because the renewal convolution \eqref{eq:renewal} only applies for $t \geq 2$. We use
\begin{equation}\label{eq:i-init}
I_{i1}  \sim
\mathrm{Lognormal}\left(\log I^{\mathrm{init}}_i -
\tfrac{1}{2 I^{\mathrm{init}}_i},
{I^{\mathrm{init}}_i}^{-1/2}\right),
\end{equation}
where the anchor $I^{\mathrm{init}}_i$ allocates a city wide weekly RSV clinical encounter count $C^{\mathrm{tot}}$ proportionally to plant populations, with a floor of 10 cases to avoid degenerate variances at small plants:
\begin{equation}\label{eq:i-init-anchor}
I^{\mathrm{init}}_i
 =
\max\left(\mathrm{round}\left(
\frac{N_i}{\sum_{i'} N_{i'}} C^{\mathrm{tot}}\right),10\right).
\end{equation}
The value $C^{\mathrm{tot}} = 324$ cases per week was computed as the mean of the five week trailing average of RSV attributed healthcare visits across the 9 Houston Health Department weekly respiratory disease surveillance reports of the 2024--2025 season that report this quantity directly \citep{houstonhealth2025respiratory}. The per report compilation used for this calculation is available in the project's \blind{\href{https://github.com/hou-wastewater-epi-org/Inferring-Spatial-Transmission-Dynamics-of-RSV-Across-Houston-Wastewater-Treatment-Plants}{GitHub repository}}{GitHub repository (link withheld for anonymized review)}. The anchor enters the model only through the initial week distribution \eqref{eq:i-init}; subsequent weeks are governed entirely by the renewal equation \eqref{eq:renewal} and the data, so the anchor's role is to keep $I_{i1}$ on a plausible scale rather than to drive the trajectory.

\subsubsection{Viral load process}\label{sec:viral-load}

Following the single plant formulation, latent infections at plant $i$ are mapped to expected viral load $\psi_{it}$ via the shedding kernel and a plant specific multiplicative scale $\beta_i > 0$:
\begin{equation}\label{eq:psi}
\psi_{it}  =  \beta_i \sum_{\tau=0}^{S} s_\tau I_{i, t-\tau}.
\end{equation}

As in \citet{palacio2026inferring}, the viral loads are modeled by a log-normal likelihood:
\begin{equation}\label{eq:obs-lik}
 y_{it}
 \sim
\mathrm{Lognormal}\left(\log(\psi_{it} + 10^{-6}) - \tfrac{1}{2}\sigma_{i}^{2}, \sigma_{i}\right), 
\end{equation}
where $\sigma_i > 0$ is the plant-specific log-scale observation noise standard deviation, estimated jointly with the remaining model parameters under a weakly informative prior specified in Section \ref{sec:computation}. The location parameter is constructed so that the natural scale mean of the lognormal coincides with the expected viral load: the $-\tfrac{1}{2}\sigma_{i}^{2}$ correction exactly cancels the $+\tfrac{1}{2}\sigma_{i}^{2}$ inflation in the lognormal mean formula, giving $\mathbb{E}[y_{it} \mid \psi_{it}] = \psi_{it} + 10^{-6} \approx \psi_{it}$. The small constant $10^{-6}$ keeps the logarithm finite when $\psi_{it}$ is near zero. Weeks without sampling contribute no term to the likelihood.

\subsection{Priors and posterior inference}\label{sec:computation}

The  Bayesian renewal model has three top-level free parameters: the spatial coupling parameter $\rho$, the plant-specific scales $\beta_1, \dots, \beta_m$, and the plant-specific log-scale observation noise 
standard deviations $\sigma_1, \dots, \sigma_m$. We assign weakly informative priors,
\begin{align}
\log \beta_i &\sim \mathcal{N}(0, 4^{2}), 
& i &= 1, \dots, m, \label{eq:beta-prior} \\
\sigma_i &\sim \mathcal{N}^{+}(0, 2^{2}), 
& i &= 1, \dots, m, \label{eq:sigma-prior} \\
\rho &\sim \mathrm{Beta}(2, 2), & & \label{eq:rho-prior}
\end{align}
together with the implicit priors on the latent growth rate $\lambda_{it}$ and the latent incidence $I_{it}$ given by the recursions of Section \ref{sec:modelspec}. The $\mathrm{Beta}$ prior on $\rho$ is symmetric and centered at $0.5$, weakly favoring intermediate values of spatial coupling without ruling out the boundary cases. The prior on $\log\beta_i$ spans several orders of magnitude on the natural scale. Finally, the half-normal prior on $\sigma_i$ places most of its mass on modest  observational noise while permitting larger values if warranted by the  data.

Posterior inference is performed with Hamiltonian Monte Carlo via the No-U-Turn Sampler \citep{hoffman2014nouturn}. We run three chains with 2000 warm-up iterations and 8000 post-warm-up iterations per chain.

\subsection{Two-stage strategy}\label{sec:ssm}

Following \citet{palacio2026inferring} and \citet{ensor2026understanding}, we  consider a modular two-stage strategy that extracts the underlying viral load  signal from noisy wastewater measurements and uses it as input to the  Bayesian renewal model of Sections~\ref{sec:modelspec}--\ref{sec:computation}. The general idea is to separate complex estimation tasks into sequential  stages. This modular structure allows us to propagate uncertainty while simplifying model estimation, and provides a natural mechanism to handle missing observations. 

For Stage 1, we adopt the first-differences-twice state-space specification of \citet{ensor2026understanding}. Let $x_{it}$ denote the latent log 10 viral load at plant $i$ in week $t$. The state and observation equations are

\begin{align}
x_{it} &= 2 x_{i, t-1} - x_{i, t-2} + w_{it},
  & w_{it} &\sim \mathcal{N}(0, \sigma_{w, i}^{2}), \label{eq:ssm-state}\\
\log_{10} y_{it} &= x_{it} + v_{it},
  & v_{it} &\sim \mathcal{N}(0, \sigma_{v, i}^{2}). \label{eq:ssm-obs}
\end{align} 
The parameters $(\sigma_{v, i}, \sigma_{w, i})$ are estimated by maximum likelihood with the Kalman filter via the \texttt{MARSS} R package \citep{holmes2012marss}, yielding the filtered state mean $\widehat{x}_{i,t \mid t}$ and variance $\widehat{P}_{i,t \mid t}$.

For Stage 2, the log-scale filter mean $\widehat{x}_{i,t \mid t}$ is  converted to the natural (B gc/day) scale to serve as the fixed observation  input in the likelihood \eqref{eq:obs-lik}, while $\sqrt{\widehat{P}_{i,t \mid t}}$  is rescaled to the natural-log scale to serve as the corresponding scale  parameter. This substitutes the viral load observations and the  jointly-estimated $\sigma_i$ of Section \ref{sec:viral-load}. The priors  on $\log \beta_i$ and $\rho$ carry over unchanged; the prior on $\sigma_i$  no longer applies.

\section{Results}\label{sec:results}

We fit the Bayesian renewal model under two configurations of the  observation input, both with a first-order Gaussian random walk on the  growth rate $\lambda_{it}$ (equation \eqref{eq:lambda-walk}). In the  Data variant, the likelihood \eqref{eq:obs-lik} is applied to the viral  loads $y_{it}$ and $\sigma_i$ is estimated jointly with the latent  process. In the Filter variant, instead of the viral load observations  and $\sigma_i$, we use the outputs of the Stage-1 state-space filter of  Section \ref{sec:ssm}. Each variant is fit twice: once with $\rho$  estimated jointly under the prior \eqref{eq:rho-prior}, and once with $\rho = 0$ as a spatially independent baseline, yielding four fits in total. All fits use the prior \eqref{eq:beta-prior} on each $\beta_i$, the same PEHD-derived spatial weight  matrix $\W$, and the same HMC settings of Section \ref{sec:computation}. Results from a second-order extension of the latent random walk are  reported in Appendix \ref{app:ar2}.

\subsection{Convergence diagnostics}\label{sec:results-conv}

All four fits achieve clean convergence at the sampler settings of Section \ref{sec:computation}: no divergent transitions, no maximum treedepth hits, and the potential scale reduction factor $\widehat R \leq 1.01$ across every monitored parameter and plant. Full per-parameter and per-plant diagnostics are reported in Appendix \ref{app:diagnostics}.

\subsection{Posterior of the spatial coupling parameter}\label{sec:results-rho}

The posterior for $\rho$ is concentrated well above zero under both variants but at distinct levels. The Filter variant yields posterior mean $0.540$ with standard deviation $0.029$ and 95\% credible interval (CrI) $[0.483, 0.597]$, while the Data variant shifts upward to mean $0.757$, standard deviation $0.038$, and 95\% CrI $[0.683, 0.833]$. Both posteriors place essentially no mass below $\rho = 0.4$, ruling out the independent plant regime at high posterior probability; neither reaches the full mixing boundary $\rho = 1$.

The shift between the two variants reflects the role of the observation noise input. When $\sigma_{i}$ is estimated jointly with the latent process, more of the observed week-to-week variation is attributable to the latent epidemiological dynamics, and the model borrows more strongly from neighboring WWTPs through the spatial mixing term to explain it. The Filter variant treats some of the observation variability as known from the Stage-1 filter, leaving less residual variation for the spatial coupling to absorb. Despite this difference, the two variants agree that transmission across WWTP boundaries is a substantial driver of plant-level RSV dynamics that the independent-plant model cannot capture.

\subsection{Plant-specific scales}\label{sec:results-beta}

Table \ref{tab:beta-estimates} reports posterior means, standard deviations, and 95\% CrIs for the plant-specific scale parameters $\beta_i$ under both variants. Plant-level scales span more than an order of magnitude, from $\widehat\beta_i$ below 1 at the smallest WWTPs (WWTP5 and WWTP15, $\sim$10--20K residents) to $\widehat\beta_i \approx 28$ at WWTP1 (578K residents), mirroring the heterogeneity of WWTP populations. The two variants differ by less than $10\%$ at every plant, and the 95\% CrIs overlap substantially for every plant. The differences  between variants in $\widehat\beta_i$ are small compared with the  variation across plants.

\begin{table}[H]
\centering
\caption{Posterior mean, standard deviation, and 95\% credible
interval of the plant specific scale parameter $\beta_i$ (in B gc/day per infection) under the Filter and Data variants. Standard deviations are computed from the posterior draws; CrIs are equal tailed. Values are from the $\rho$-estimated fits.}
\label{tab:beta-estimates}
\footnotesize
\begin{tabular}{lccc ccc}
\toprule
& \multicolumn{3}{c}{Filter} & \multicolumn{3}{c}{Data} \\
\cmidrule(lr){2-4} \cmidrule(lr){5-7}
 & Mean & SD & 95\% CrI & Mean & SD & 95\% CrI \\
\midrule
WWTP1 & 27.52 & 1.35 & $[24.93,\ 30.23]$ & 26.13 & 1.47 & $[23.44,\ 29.20]$ \\
WWTP2 & 4.09 & 0.19 & $[3.72,\ 4.48]$ & 3.78 & 0.20 & $[3.41,\ 4.19]$ \\
WWTP3 & 4.45 & 0.24 & $[4.00,\ 4.92]$ & 4.56 & 0.28 & $[4.04,\ 5.13]$ \\
WWTP4 & 2.37 & 0.13 & $[2.13,\ 2.62]$ & 2.45 & 0.16 & $[2.15,\ 2.78]$ \\
WWTP5 & 0.94 & 0.05 & $[0.84,\ 1.04]$ & 0.96 & 0.06 & $[0.84,\ 1.09]$ \\
WWTP6 & 3.18 & 0.17 & $[2.86,\ 3.51]$ & 3.06 & 0.17 & $[2.74,\ 3.42]$ \\
WWTP7 & 5.23 & 0.29 & $[4.68,\ 5.81]$ & 5.34 & 0.30 & $[4.78,\ 5.97]$ \\
WWTP8 & 10.03 & 0.47 & $[9.12,\ 10.97]$ & 9.40 & 0.46 & $[8.52,\ 10.34]$ \\
WWTP9 & 2.53 & 0.13 & $[2.28,\ 2.80]$ & 2.72 & 0.18 & $[2.40,\ 3.09]$ \\
WWTP10 & 4.97 & 0.24 & $[4.51,\ 5.45]$ & 5.09 & 0.26 & $[4.60,\ 5.63]$ \\
WWTP11 & 3.74 & 0.22 & $[3.32,\ 4.19]$ & 4.14 & 0.27 & $[3.64,\ 4.71]$ \\
WWTP12 & 8.11 & 0.46 & $[7.23,\ 9.02]$ & 8.73 & 0.53 & $[7.76,\ 9.83]$ \\
WWTP13 & 4.45 & 0.20 & $[4.06,\ 4.85]$ & 4.26 & 0.23 & $[3.82,\ 4.74]$ \\
WWTP14 & 12.76 & 0.58 & $[11.64,\ 13.91]$ & 12.93 & 0.60 & $[11.80,\ 14.14]$ \\
WWTP15 & 0.76 & 0.05 & $[0.66,\ 0.86]$ & 0.82 & 0.08 & $[0.69,\ 0.99]$ \\
\bottomrule
\end{tabular}
\end{table}

\subsection{Effective reproduction number and incidence trajectories}\label{sec:results-trajectories}

\begin{figure}[H]
\centering
\includegraphics[width=\linewidth]{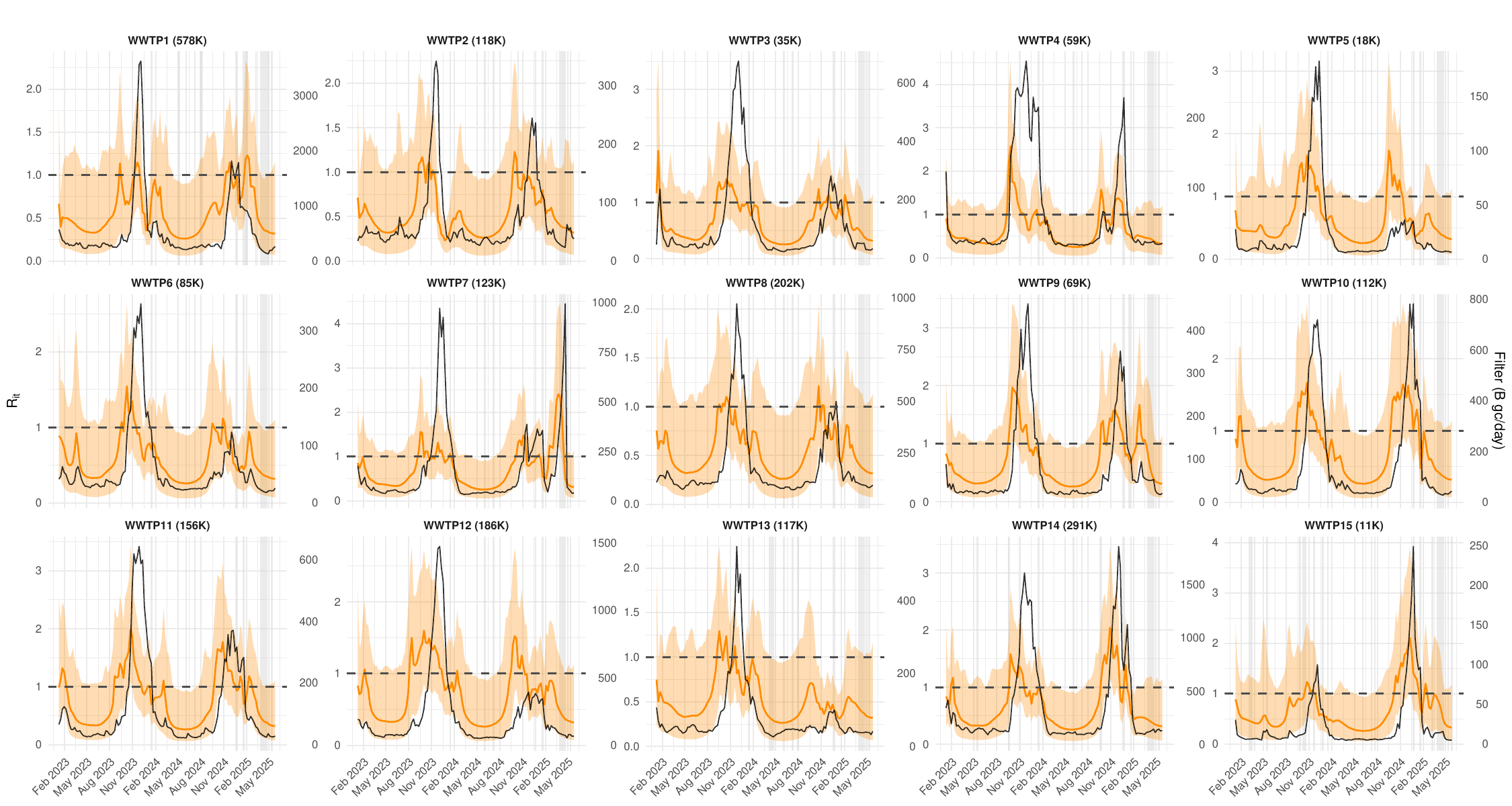}
\caption{Posterior $R_{it}$ under the Filter variant with $\rho$ estimated, shown by week (x-axis) for the 15 Houston plants (one panel per plant). The y-axis gives $R_{it}$. Dark line: posterior median. Light band: 95\% CrI. The dashed horizontal line at $R_{it} = 1$ marks the epidemic threshold. Vertical gray bars mark weeks with no sample.}
\label{fig:Rt-ssm-est}
\end{figure}

\begin{figure}[H]
\centering
\includegraphics[width=\linewidth]{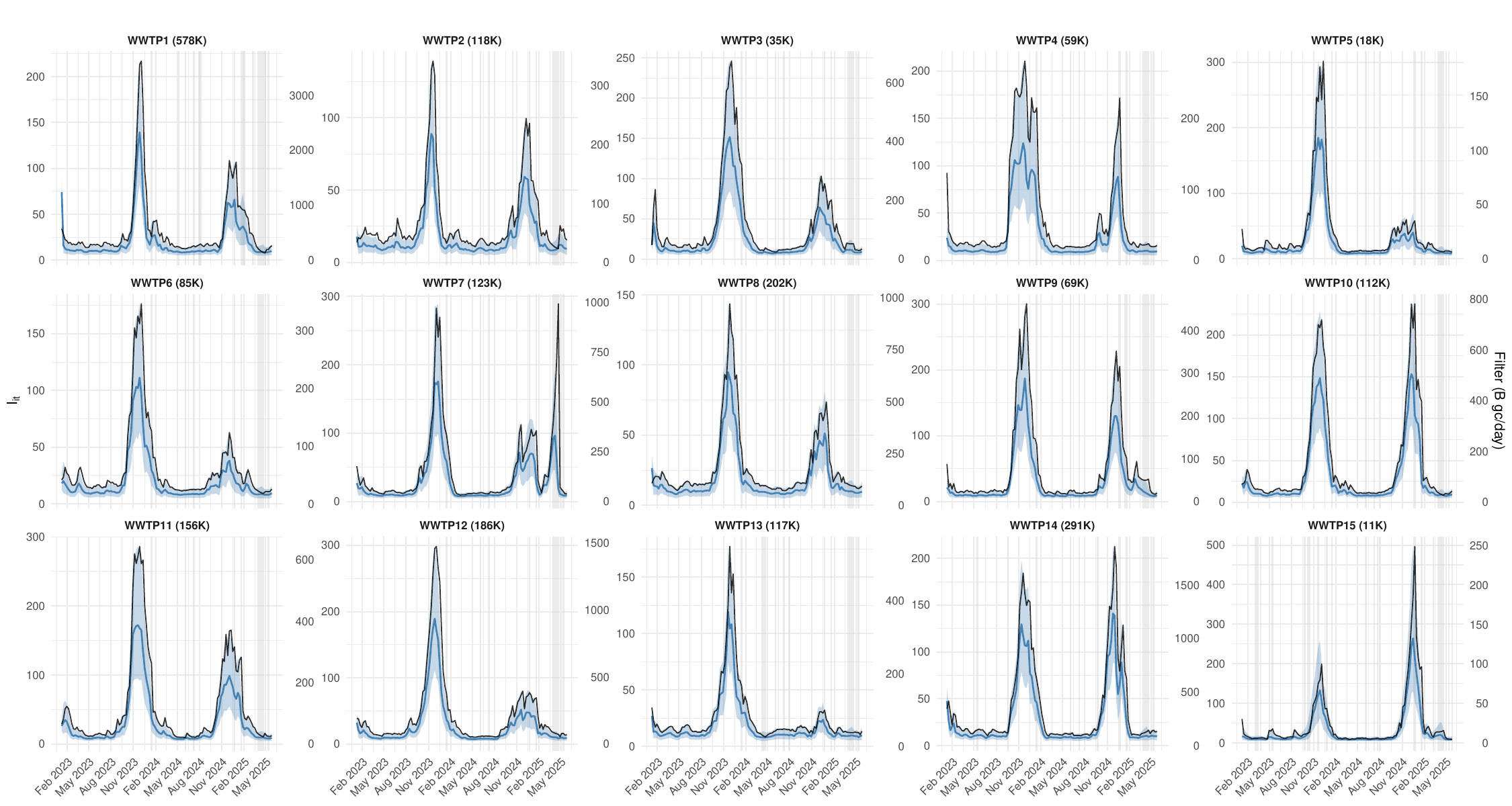}
\caption{Posterior $I_{it}$ under the Filter variant with $\rho$ estimated, shown by week (x-axis) for the 15 Houston plants (one panel per plant). The left y-axis gives $I_{it}$. The observed viral load $y_{it}$ (B gc/day) is overlaid in black on the right y-axis of each panel. Vertical gray bars mark weeks with no sample.}
\label{fig:It-ssm-est}
\end{figure}

\begin{figure}[H]
\centering
\includegraphics[width=\linewidth]{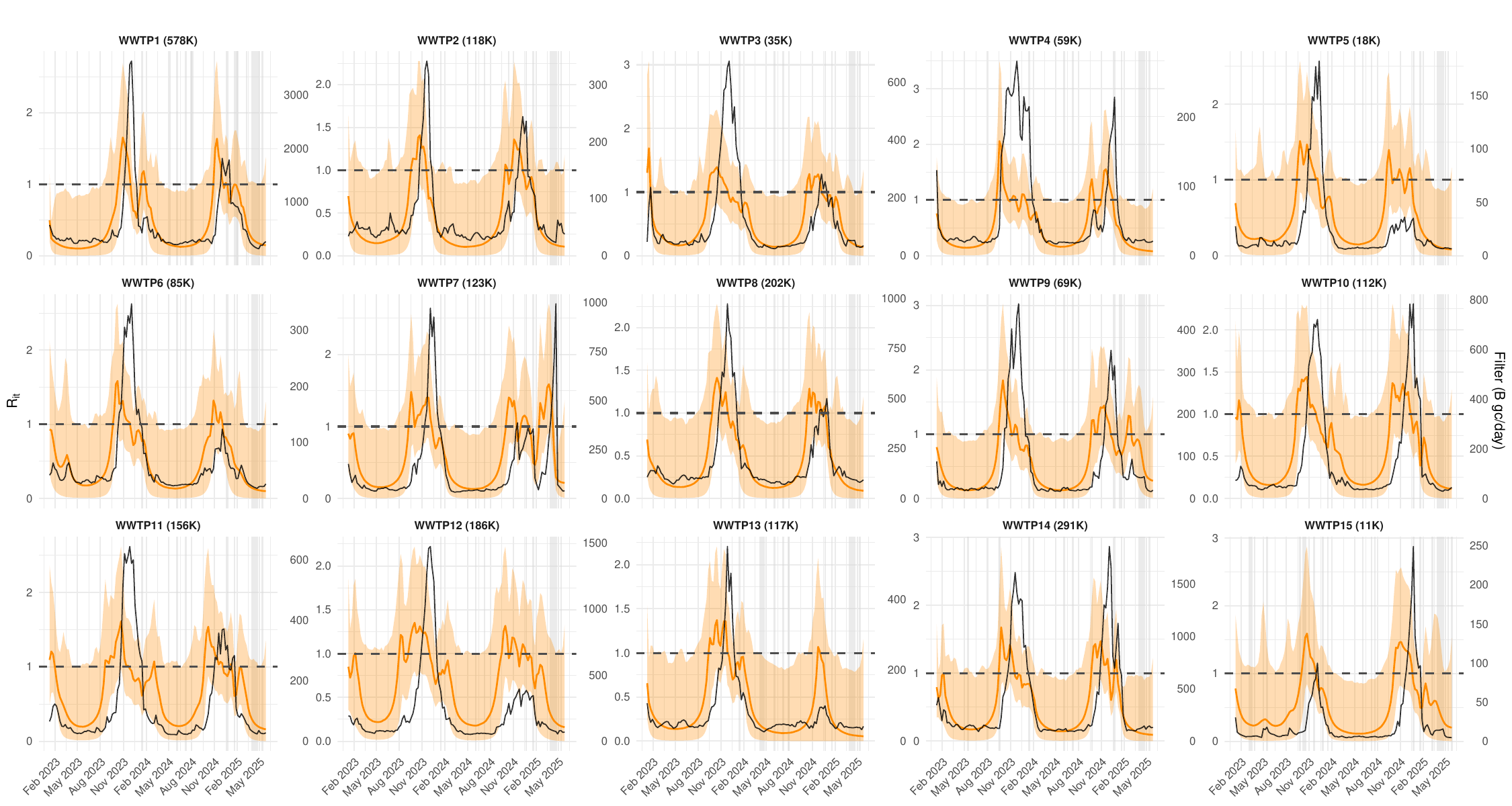}
\caption{Posterior $R_{it}$ under the Filter variant
with $\rho = 0$.}
\label{fig:Rt-ssm-fix}
\end{figure}

\begin{figure}[H]
\centering
\includegraphics[width=\linewidth]{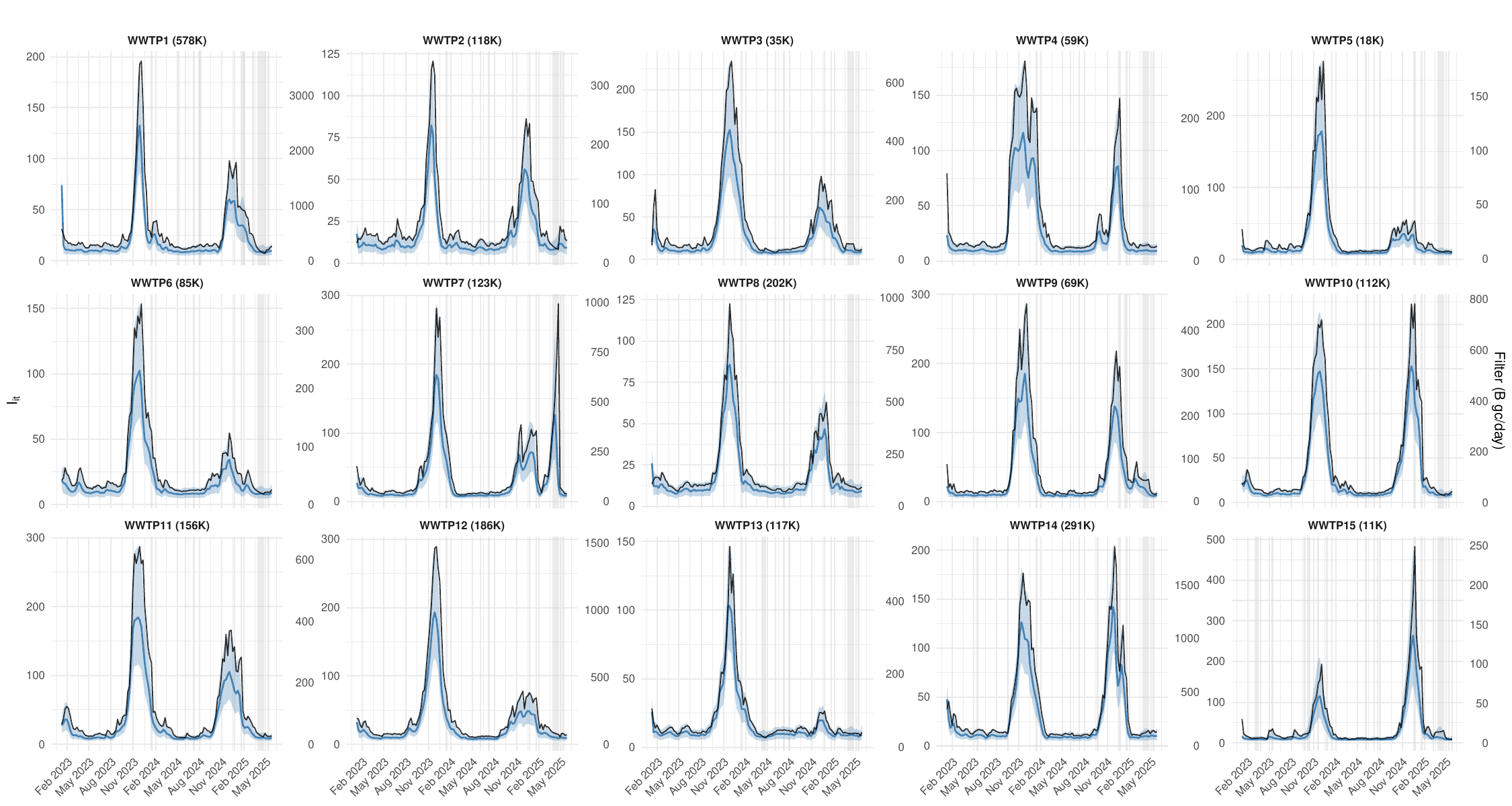}
\caption{Posterior $I_{it}$ under the Filter  variant
with $\rho = 0$.}
\label{fig:It-ssm-fix}
\end{figure}

\begin{figure}[H]
\centering
\includegraphics[width=\linewidth]{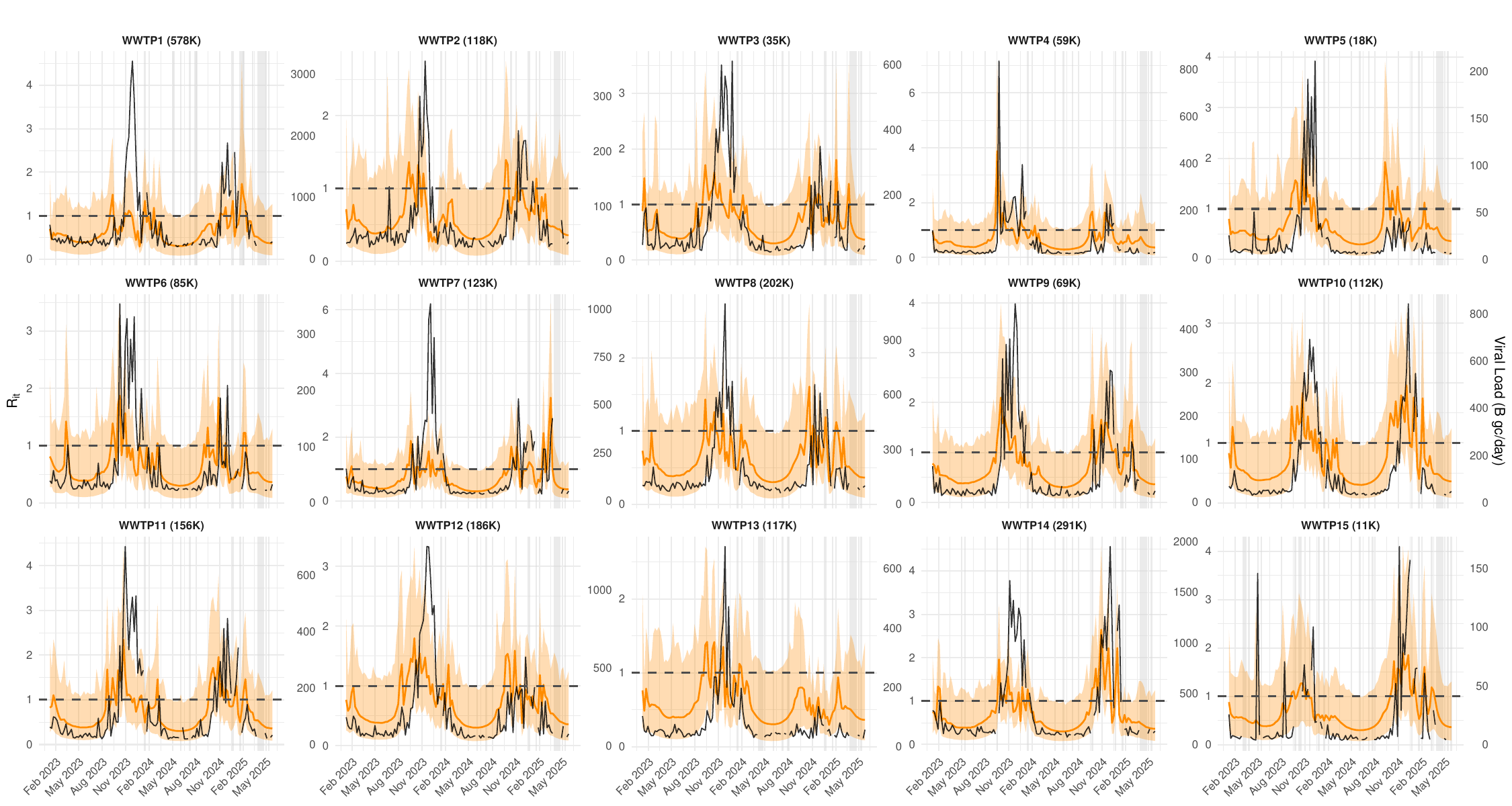}
\caption{Posterior $R_{it}$ under the Data  variant
with $\rho$ estimated.}
\label{fig:Rt-direct-est}
\end{figure}

\begin{figure}[H]
\centering
\includegraphics[width=\linewidth]{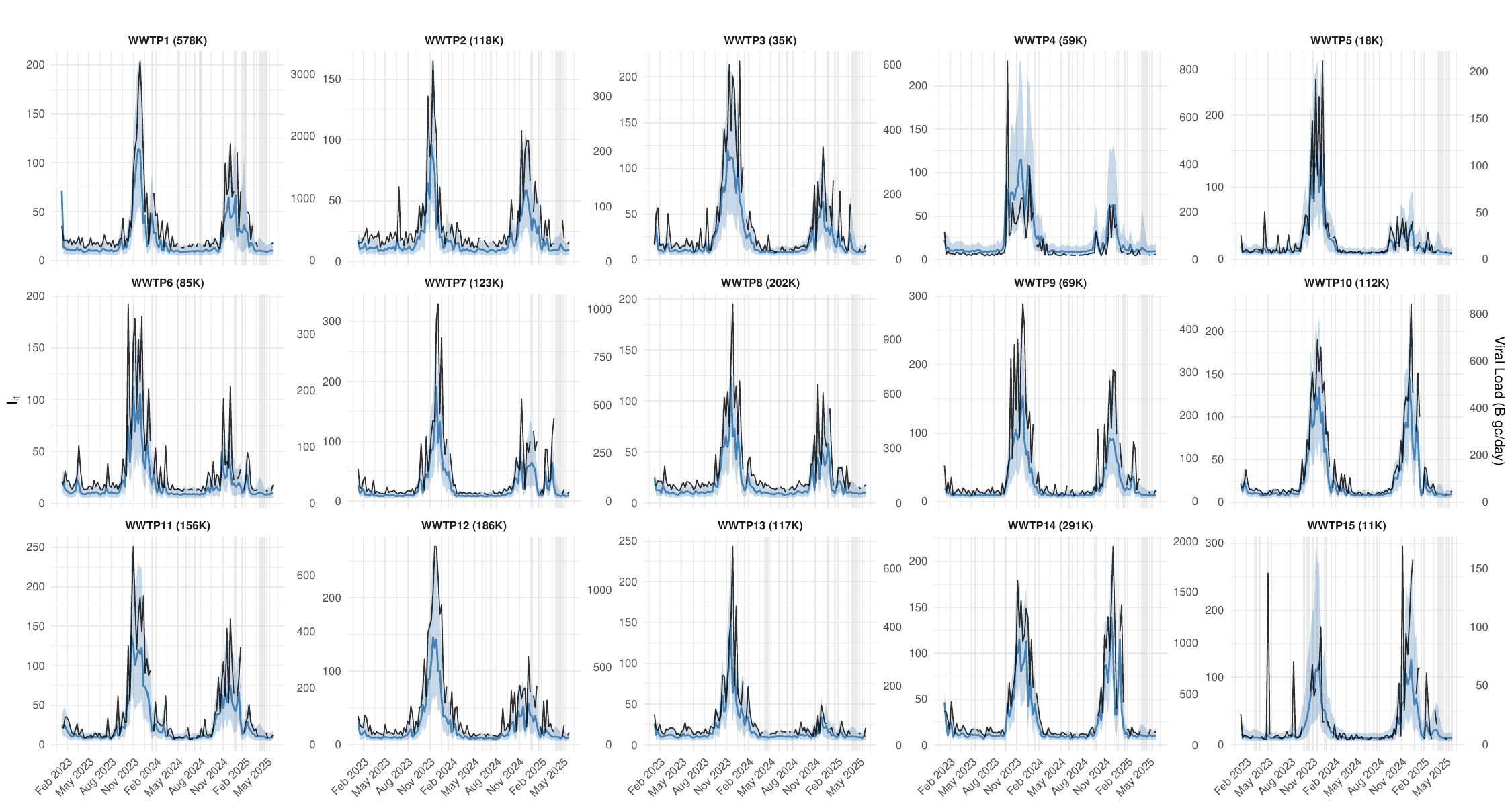}
\caption{Posterior $I_{it}$ under the Data  variant
with $\rho$ estimated.}
\label{fig:It-direct-est}
\end{figure}

\begin{figure}[H]
\centering
\includegraphics[width=\linewidth]{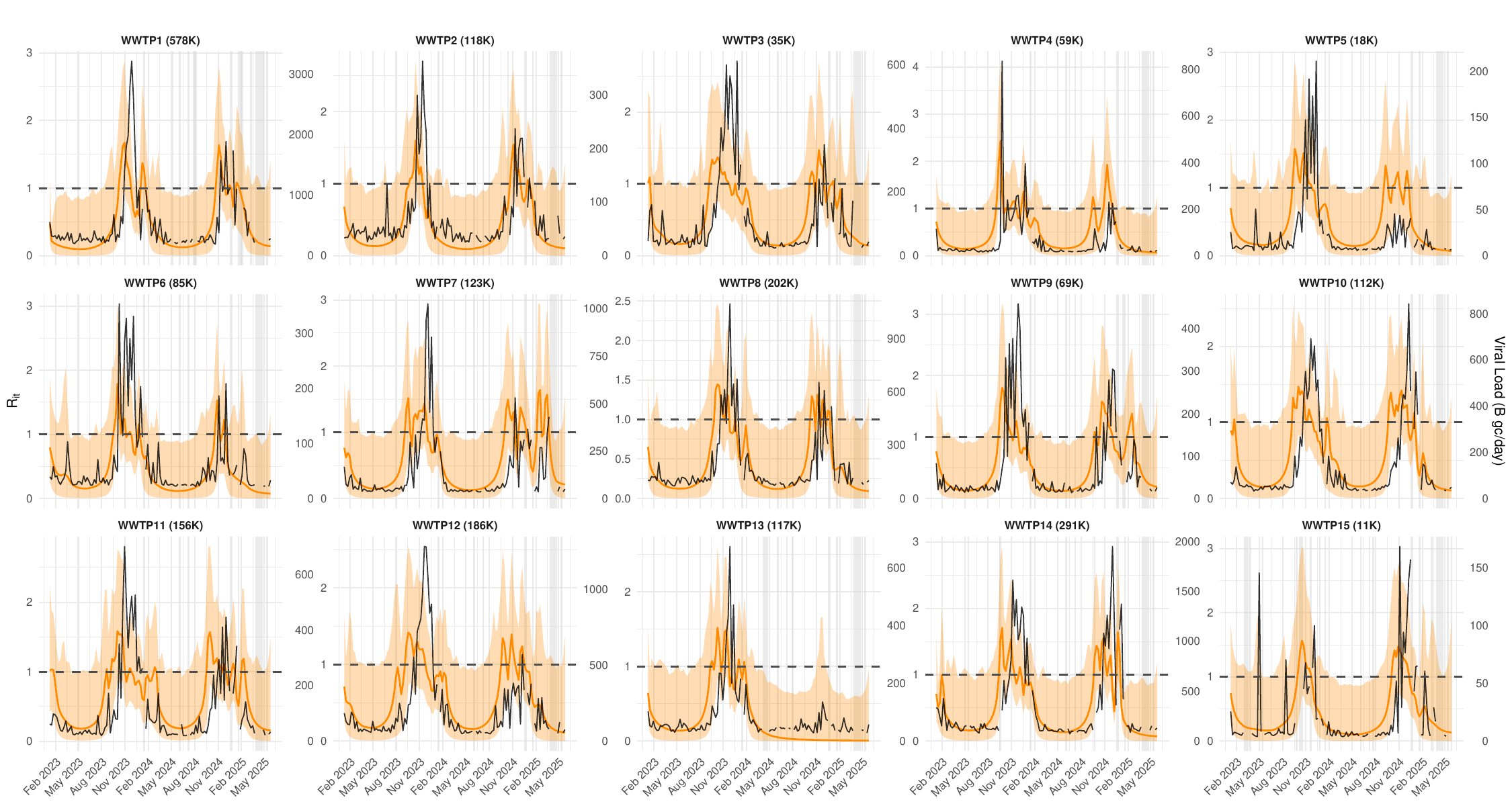}
\caption{Posterior $R_{it}$ under the Data  variant
with $\rho = 0$.}
\label{fig:Rt-direct-fix}
\end{figure}

\begin{figure}[H]
\centering
\includegraphics[width=\linewidth]{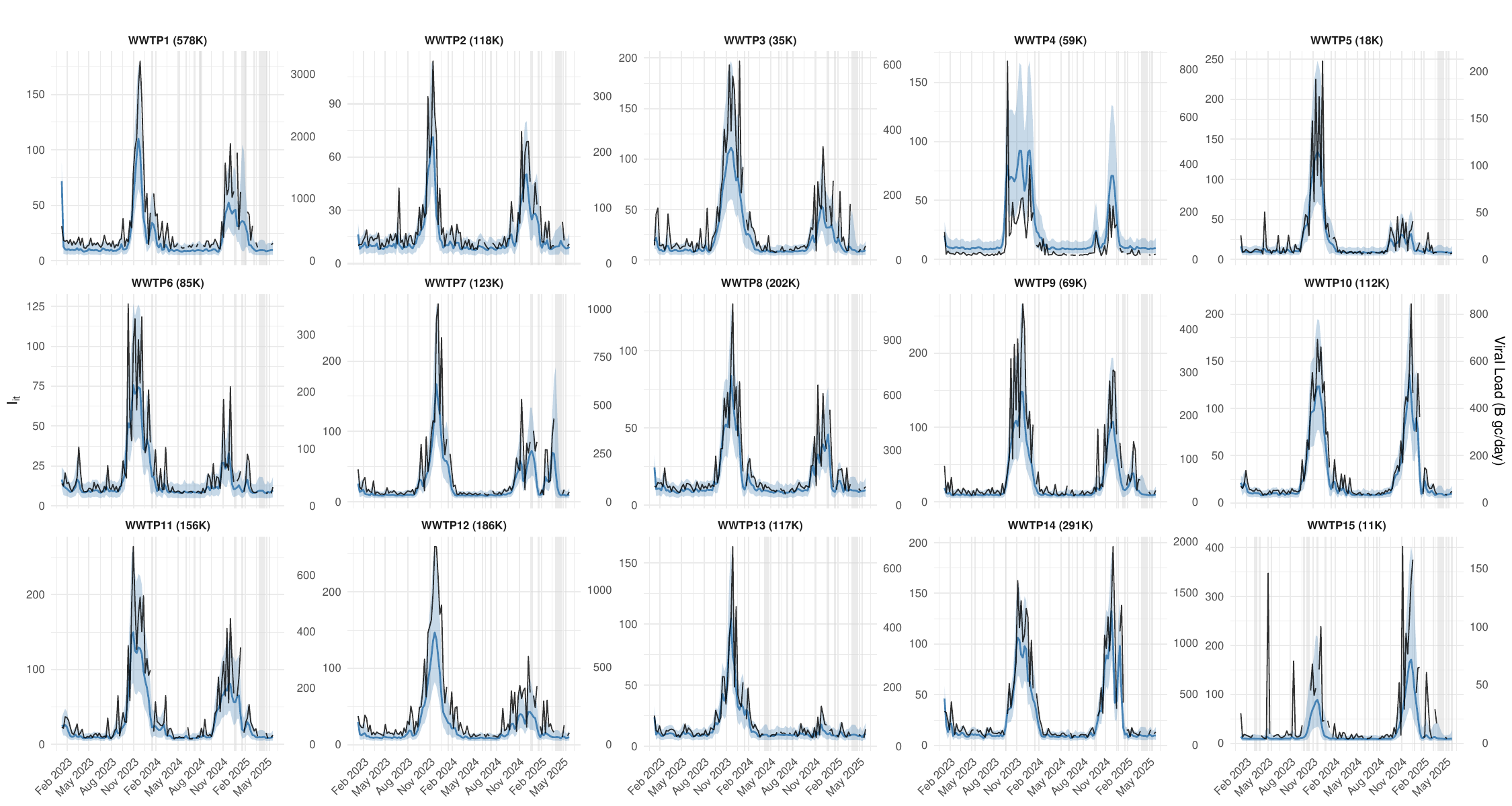}
\caption{Posterior $I_{it}$ under the Data  variant
with $\rho = 0$.}
\label{fig:It-direct-fix}
\end{figure}

Figures \ref{fig:Rt-ssm-est}--\ref{fig:It-direct-fix} display trajectories of the posterior mean and 95\% credible band of the effective reproduction number $R_{it}$ and the weekly incidence $I_{it}$ for all 15 plants under each of the four configurations. Following the two-stage strategy of \citet{palacio2026inferring}, we treat the Filter variant with $\rho$ estimated (Figures \ref{fig:Rt-ssm-est} and \ref{fig:It-ssm-est}) as the primary configuration; the remaining three variants are reported for direct comparison. In each panel, the black overlay shows the input to the corresponding variant. The filter estimates for the Filter variant and the viral loads for the Data variant. The vertical gray bars mark weeks with no sample (also shown in the Filter variant panels for comparability). 

The seasonal RSV signal is visible across all plants and all variants: all 15 plants cross $R_{it} = 1$ from below in late summer, remain above 1 through the fall wave, and return to values below 1 by January. Peaks are broadly synchronized across the network, consistent with the substantial spatial coupling implied by $\rho \approx 0.540$ (Filter) and $\rho \approx 0.757$ (Data). The Filter variant yields narrower credible bands than the Data variant, since the Stage-1 filter separates the uncertainty in the underlying trend from that due to sampling and measurement error, so Stage-2 receives the filtered signal and its week-specific variance rather than the raw viral load. Some variability in the peaks of $R_{it}$ remains, reflecting differences in the intensity of local transmission. The inferred incidence $I_{it}$ tracks the observed wastewater signal closely under all four configurations. 

A notable case is WWTP15, where the viral load contains two isolated spikes early in  the series that do not fit the wave pattern seen across the network.  The model does not follow these spikes in either $R_{it}$ or $I_{it}$  under any of the four configurations, correctly identifying them as  measurement anomalies rather than sustained transmission. This  robustness reflects the joint action of the components in the model: the random walk on $\lambda_{it}$ penalizes abrupt  week-to-week changes, and when spatial coupling is active, the low signal from neighboring plants pulls the estimates back toward baseline. Notice that in the Filter variant, through Stage-1, the peaks are already attenuated before entering the model.


\begin{figure}[!htbp]
\centering
\includegraphics[width=\linewidth,height=0.45\textheight,keepaspectratio]{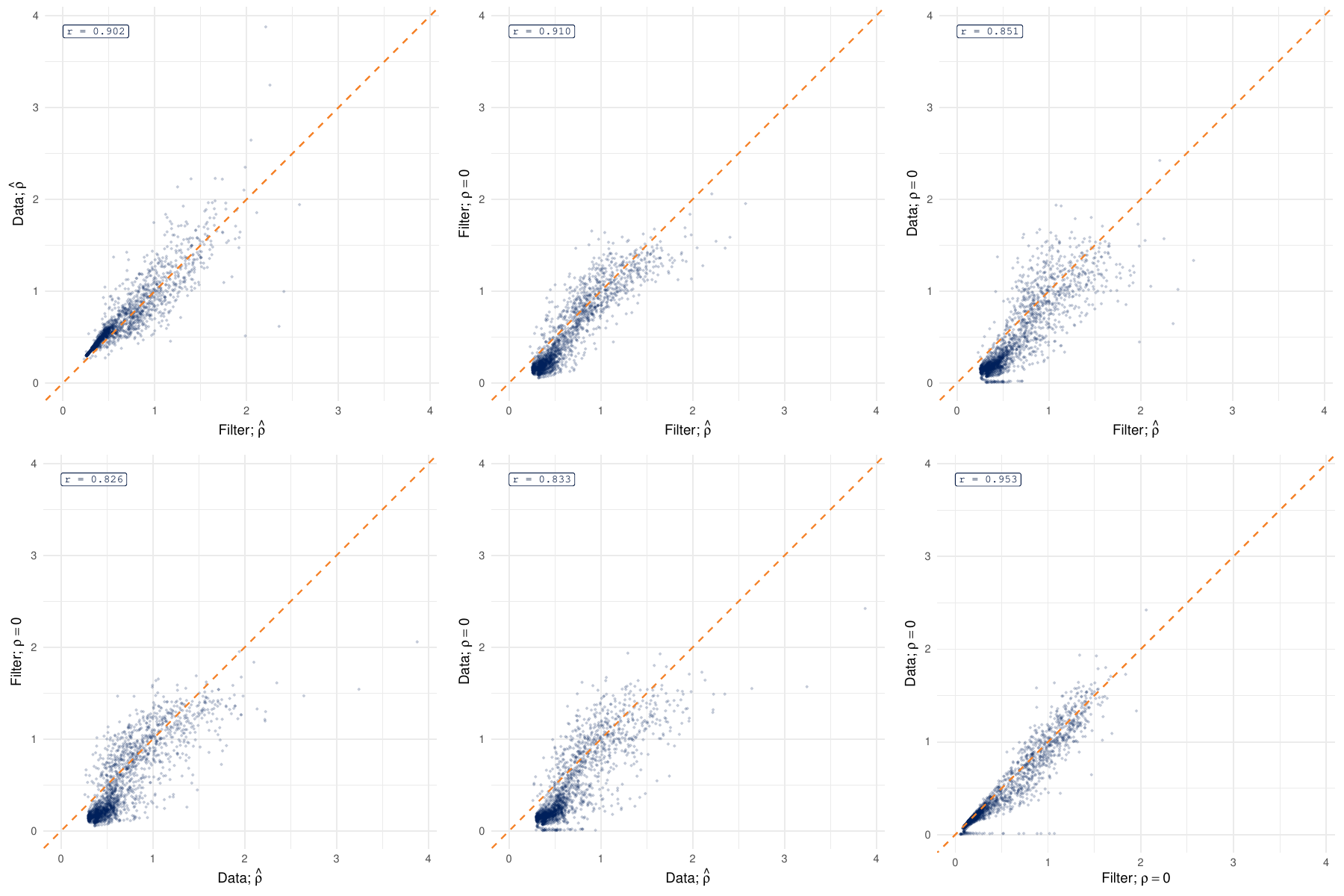}
\caption{Pairwise comparison of plant week posterior medians of
$R_{it}$ across the four fits. Each panel shows one pair, with the identity (dashed) for reference. Pearson correlation is reported in each panel.}
\label{fig:contrast-Rt}
\end{figure}

\begin{figure}[!htbp]
\centering
\includegraphics[width=\linewidth,height=0.45\textheight,keepaspectratio]{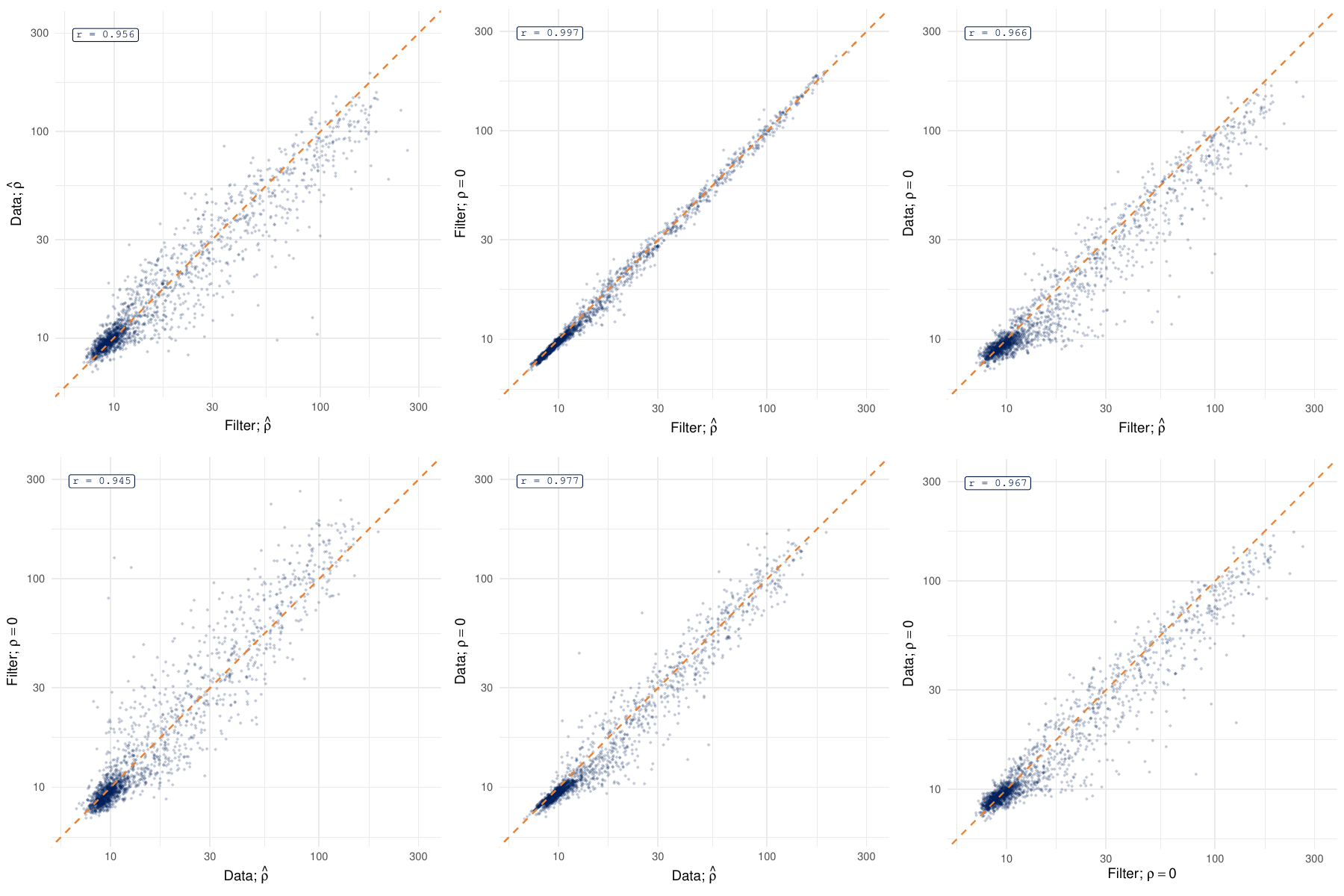}
\caption{Pairwise comparison of plant week posterior medians of
$I_{it}$ on $\log_{10}$ scale across the four fits. Tight alignment along the identity indicates that the inferred latent incidence is robust to the choice of variant and of $\rho$.}
\label{fig:contrast-It}
\end{figure}

Pointwise agreement between fits is quantified in Figures \ref{fig:contrast-Rt} and \ref{fig:contrast-It}, which plot posterior medians for every pair of the four fits. Two patterns stand out. First, $R_{it}$ correlates strongly across fits but with non-trivial scatter: within the Filter variant, the $\rho$ estimated and $\rho = 0$ fits correlate at Pearson $r = 0.91$; within the Data variant, at $r = 0.83$; across variants with $\rho$ estimated, at $r = 0.90$. Second, the $I_{it}$ scatters are tightly aligned with the identity for every pair: Pearson correlations on $\log_{10}$ scale range from $r = 0.95$ (Data $\hat\rho$ vs Filter $\rho = 0$) to $r = 0.997$ (Filter $\hat\rho$ vs Filter $\rho = 0$). The latent incidence is therefore robust to the choice of variant and of $\rho$; $R_{it}$ shows modest variation, most visibly between the $\rho$-estimated and $\rho = 0$ fits. These differences stem from the model's architecture. $I_{it}$ enters the observation likelihood through $\psi_{it}$ and is therefore more strongly influenced by the observed viral load, whereas $R_{it}$ is not part of the likelihood directly and can shift more freely. The remaining variation in $R_{it}$ across configurations is driven mainly by the spatial coupling term and the choice of input.

\subsection{In-sample predictive fit via CRPS}\label{sec:results-crps}

We assess in-sample model calibration via the mean continuous ranked  probability score (CRPS) \citep{gneiting2007strictly}, a proper scoring rule that jointly measures the sharpness and calibration of a predictive distribution against a scalar observation, with lower values indicating better fit. For each plant $i$ and week $t$ with an observation, we draw posterior predictive samples of the wastewater viral load (B gc/day) from each fitted model, compute the CRPS against the observation $y_{it}$ using \texttt{crps\_sample()} from the \texttt{scoringRules} R package \citep{jordan2019scoringrules}, then report the mean across weeks for each plant. The evaluation is in-sample and scored against $y_{it}$, the only directly observable quantity in the pipeline.

Table~\ref{tab:crps-by-plant} reports per-plant mean CRPS for the Data  model with $\rho$ estimated and with $\rho = 0$, against the EpiSewer single-plant baseline. Because CRPS is reported in the units of the observation (B gc/day) and viral load scales vary by more than an order of magnitude across plants, values are comparable only within a across models, not across plants. Both Data variants outperform EpiSewer at every plant, with per-plant reductions ranging from $19\%$ to $49\%$ under the Data variant with $\hat\rho$ and from $18\%$ to $39\%$ under $\rho = 0$. When we compare the two variants directly, the Data variant with $\hat\rho$ achieves lower CRPS at 14 of the 15 plants (the exception is at WWTP15, the plant with the anomalous spikes noted in Section~\ref{sec:results-trajectories}). Most of the improvement over EpiSewer is captured by the $\rho = 0$ variant, indicating that the bulk of the gain comes from the underlying renewal model specification (Section~\ref{sec:modelspec}); estimating $\rho$ jointly provides an additional contribution to every plant.

\begin{table}[H]
\centering
\caption{Per-plant mean in-sample CRPS (B gc/day) against the observed  viral load $y_{it}$. EpiSewer \citep{lison2025episewer} is the single-plant baseline. Data $\rho = 0$ and Data 
$\hat\rho$ are the proposed models with $\rho$ fixed and estimated. Lower is better.}
\label{tab:crps-by-plant}
\footnotesize
\begin{tabular}{l c c c c}
\toprule
 & & \multicolumn{3}{c}{Mean CRPS (B gc/day)} \\
\cmidrule(lr){3-5}
Plant & $n_{\text{obs}}$ & EpiSewer & Data $\rho = 0$ & Data $\hat\rho$ \\
\midrule
WWTP1  & 108 & 122.7 & 83.8  & 73.0 \\
WWTP2  & 114 &  14.7 & 10.0  &  9.0 \\
WWTP3  & 115 &  26.8 & 21.2  & 18.4 \\
WWTP4  & 115 &  21.4 & 16.6  & 15.3 \\
WWTP5  & 115 &   6.7 &  5.4  &  4.9 \\
WWTP6  & 115 &  16.1 & 12.1  &  9.7 \\
WWTP7  & 115 &  38.0 & 25.6  & 21.9 \\
WWTP8  & 115 &  35.4 & 26.4  & 21.4 \\
WWTP9  & 114 &  20.5 & 15.1  & 14.9 \\
WWTP10 & 115 &  24.7 & 17.2  & 14.8 \\
WWTP11 & 115 &  28.6 & 21.3  & 20.0 \\
WWTP12 & 115 &  44.2 & 36.4  & 31.8 \\
WWTP13 & 111 &  16.6 & 12.4  & 10.9 \\
WWTP14 & 112 &  66.0 & 40.2  & 33.9 \\
WWTP15 & 103 &  10.1 &  8.0  &  8.2 \\
\bottomrule
\end{tabular}
\end{table}

\section{Discussion}
\label{sec:discussion}

We extended a single-plant Bayesian renewal model to $m = 15$ Houston wastewater treatment plants by coupling neighboring WWTPs through a population-weighted spatial mixing parameter $\rho \in [0,1]$. The posterior of $\rho$ concentrated well above zero under both variants (Filter mean $0.540$, 95\% CrI $[0.483, 0.597]$; Data mean $0.757$, 95\% CrI $[0.683, 0.833]$), implying that approximately half (Filter) to three quarters (Data) of the latent growth-rate dynamics at each plant reflects the neighbors' past rather than its own. This signal is consistent with the well-documented fact that respiratory contact networks (schools, workplaces, transit, mixed households) routinely cross sanitary-engineering WWTP boundaries \citep{held2005statistical, meyer2014powerlaw}. The shift in $\hat\rho$ between the two variants is informative: when the Stage-1 filter separates the underlying trend from the sampling and measurement error, less week-to-week variation remains for the spatial mixing term to explain, and $\hat\rho$ settles near $0.540$; when $\sigma_i$ is estimated jointly with the latent process, the spatial mechanism carries more of the structural load and $\hat\rho$ rises to $0.757$. Despite this difference, both variants agree that transmission across WWTP boundaries is a substantial driver of plant-level RSV dynamics, one that the independent-plant model ($\rho = 0$) cannot capture.

Given the robustness of $I_{it}$ and the moderate sensitivity of $R_{it}$ across configurations (Section~\ref{sec:results-trajectories}), we recommend reporting $I_{it}$ as the primary operational metric in spatial wastewater surveillance, with $R_{it}$ accompanied by appropriately wide credible intervals. These results suggest that surveillance designs informed by population-weighted distance could potentially capture regional dynamics with fewer monitoring sites, which would be especially useful for resource-constrained health departments. In addition, the Data model achieved lower in-sample CRPS than the EpiSewer single-plant baseline \citep{lison2025episewer} at every plant, with per-plant reductions of $19$--$49\%$ under $\hat\rho$ (Section~\ref{sec:results-crps}); most of the improvement comes from the underlying renewal model specification, with estimating $\rho$ jointly providing an additional contribution.

The analysis is restricted to a single pathogen (RSV) in a single city (Houston); generalization requires replication for other respiratory viruses (influenza, SARS-CoV-2) and metropolitan areas with different network geometries and demographic profiles. The random walk on $\lambda$ and the time-invariant spatial weight matrix $\W$ are convenient simplifications that may not capture nonstationary mobility patterns or season-specific mixing; time-varying $\W$ and richer latent dynamics are natural extensions. Extensions that treat $\W$ as a covariance matrix, rather than as the matrix of mixing weights used here, would need to verify that the underlying distance yields a valid covariance structure, which is not guaranteed for set-based or directional distances such as the PEHD \citep{schoenberg1938metric}. Recent work by \citet{cunhagodoy2026hausdorff} develops Hausdorff-based Gaussian processes for areal data, providing a template for the classical symmetric case; extending such a construction to directional and population-weighted variants remains an open problem. The inferred quantities $R_{it}$ and $I_{it}$ are latent, with no ground-truth observations against which to validate them directly; our in-sample CRPS evaluation targets the observed $y_{it}$, providing a basic fit check. Out-of-sample evaluations and joint modeling of wastewater with clinical encounter data (where available) are natural next steps for further validation. The framework itself is pathogen-agnostic: the generation kernel, shedding profile, and incidence anchor are the only pathogen-specific elements, and the spatial weight construction transfers directly to other contexts. As multi-pathogen wastewater surveillance expands \citep{boehm2023wastewater}, spatial Bayesian renewal models of this type provide a principled framework for extracting regional transmission dynamics across WWTP networks.

\section{Disclosure statement}\label{sec:disclosure}

No potential conflict of interest was reported by the author(s).

\section{Data Availability Statement}\label{sec:data-availability}

Code and reproducibility materials are available in the project's GitHub repository\blind{: \url{https://github.com/hou-wastewater-epi-org/Modeling-Transmission-Dynamics-Using-Wastewater-and-PWD},}{ (link withheld for anonymized review).} together with any data and supporting files that can be distributed under applicable confidentiality agreements.

\section{AI Usage Disclosure}\label{sec:ai-usage}

During the preparation of this manuscript, the corresponding author\blind{ Jose R. Palacio}{} used Claude Version Opus 4.7 as writing and computational assistants. The assistance included grammar and style correction, translation between Spanish and English, refinement of phrasing, debugging R and Stan code, and resolving \LaTeX{} and notation inconsistencies. All methodological decisions, model specifications, interpretation of results, and substantive scientific claims were made and verified by the human authors. \blind{Jose R. Palacio}{The corresponding author} reviewed and edited all AI generated content before incorporating it into the manuscript and takes full responsibility for the content of this publication.

\if1\anon
\section{Funding and Acknowledgments}\label{sec:funding}

The authors disclosed receipt of the following financial support for the research, authorship, and/or publication of this article: This work was supported by the Centers for Disease Control and Prevention (ELC ED grant no. 6NU50CK000557 01 05 and ELC CORE grant no. NU50CK000557). The authors acknowledge Houston Public Works for their contributions to the HHD WBE system. The authors would like to acknowledge the CDC National Wastewater Surveillance System (NWSS) scientific community. For more information on the Houston Wastewater Epidemiology Center of Excellence see \url{https://www.hou-wastewater-epi.org}.

\section{Notes on contributor(s)}\label{sec:contributors}

Palacio is the corresponding author, representing both intellectual leadership and implementation of the methodologies. He presented the team with a complete first draft for editorial comments. Ensor, Rice Co-PI of this project, led the technical development and draft writing. Keller consulted on developing the two-stage implementation with specific attention to uncertainty quantification. Schedler contributed to the methodology section and provided the code for the PEHD implementation. Schneider oversees the day-to-day implementation of the HHD WBE system and represents the team in national data analysis conversations. Domakonda serves as HHD manager for this project. Hopkins serves as HHD PI for this project. Stadler, Rice Co-PI of this project, oversees all aspects of laboratory analyses and brings essential expertise in WBE. Bhandari provided the census tract map and population data. All authors contributed to the scientific discussion and edited the manuscript.
\fi

\phantomsection\label{supplementary-material}
\bigskip

\begin{center}

{\large\bf SUPPLEMENTARY MATERIAL}

\end{center}

\begin{description}
\item[Code and data:]
R and Stan code to reproduce the analyses described in the article, together with the compilation of Houston Health Department surveillance reports used to compute $C^{\mathrm{tot}}$ in Section \ref{sec:incidence}. (GitHub repository\blind{}{; link withheld for anonymized review})
\end{description}

\bibliography{references}

\appendix

\section{Convergence diagnostics}\label{app:diagnostics}

\begin{figure}[H]
\centering
\includegraphics[width=\linewidth]{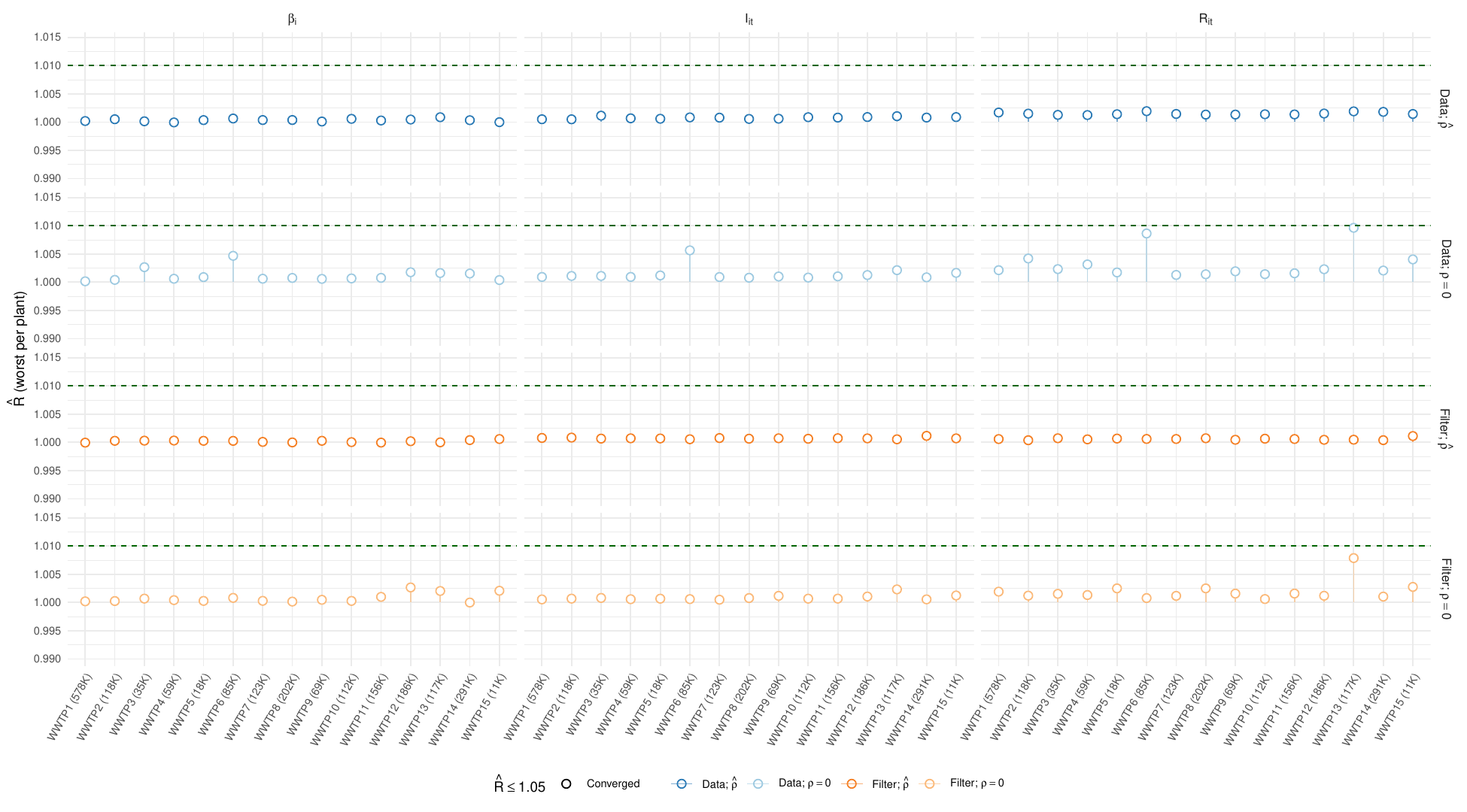}
\caption{Worst case $\widehat R$ per plant within each parameter
block ($\beta_i$, $I_{it}$, $R_{it}$), faceted by fit variant (rows) and block (columns). White points indicate plants whose worst $\widehat R$ is below the conservative threshold of $1.05$; orange would indicate plants above it. Every plant in every block satisfies $\widehat R \leq 1.01$ across all four fits.}
\label{fig:rhat-lollipop}
\end{figure}

This appendix reports the per parameter convergence diagnostics for all four fits (Filter and Data variants, each with $\hat\rho$ and $\rho = 0$). Figure \ref{fig:rhat-lollipop} reports the worst case potential scale reduction factor $\widehat R$ per plant within each parameter block ($\beta_i$, $I_{it}$, $R_{it}$), faceted by fit variant. All monitored parameters satisfy $\widehat R \leq 1.01$ across the four fits, well below the conservative threshold of $1.05$ recommended by \citet{vehtari2017practical}.

\section{Second order extension of the latent dynamics}\label{app:ar2}

In the main specification, the latent growth rate $\lambda_{it}$ follows a spatially coupled first order Gaussian random walk (equation \eqref{eq:lambda-walk}). This appendix reports a second order extension in which $\lambda_{it}$ evolves as
\begin{equation}\label{eq:lambda-walk-ar2}
\lambda_{it}  \sim  \mathcal{N}\bigl((1-\rho)(2\lambda_{i, t-1} - \lambda_{i, t-2})
+ \rho \sum_{j=1}^{m} \W_{ij}(2\lambda_{j, t-1} - \lambda_{j, t-2}),\ \sigma_{\lambda}\bigr),
\end{equation}
i.e., a Gaussian random walk on the differences $\Delta\lambda_{it} = \lambda_{it} - \lambda_{i, t-1}$ rather than on the levels. The second order process is locally smoother than the first order walk: in the absence of new shocks, the local trend in $\lambda_{it}$ persists, which is better suited to sustained acceleration or deceleration of transmission than the unit shock dynamics of equation \eqref{eq:lambda-walk}. All other components of the model (the renewal convolution, the shedding kernel, the spatial weight matrix $\W$, the priors on $\boldsymbol\beta$ and $\rho$, and the HMC settings of Section \ref{sec:computation}) are unchanged.

\begin{figure}[H]
\centering
\includegraphics[width=\linewidth]{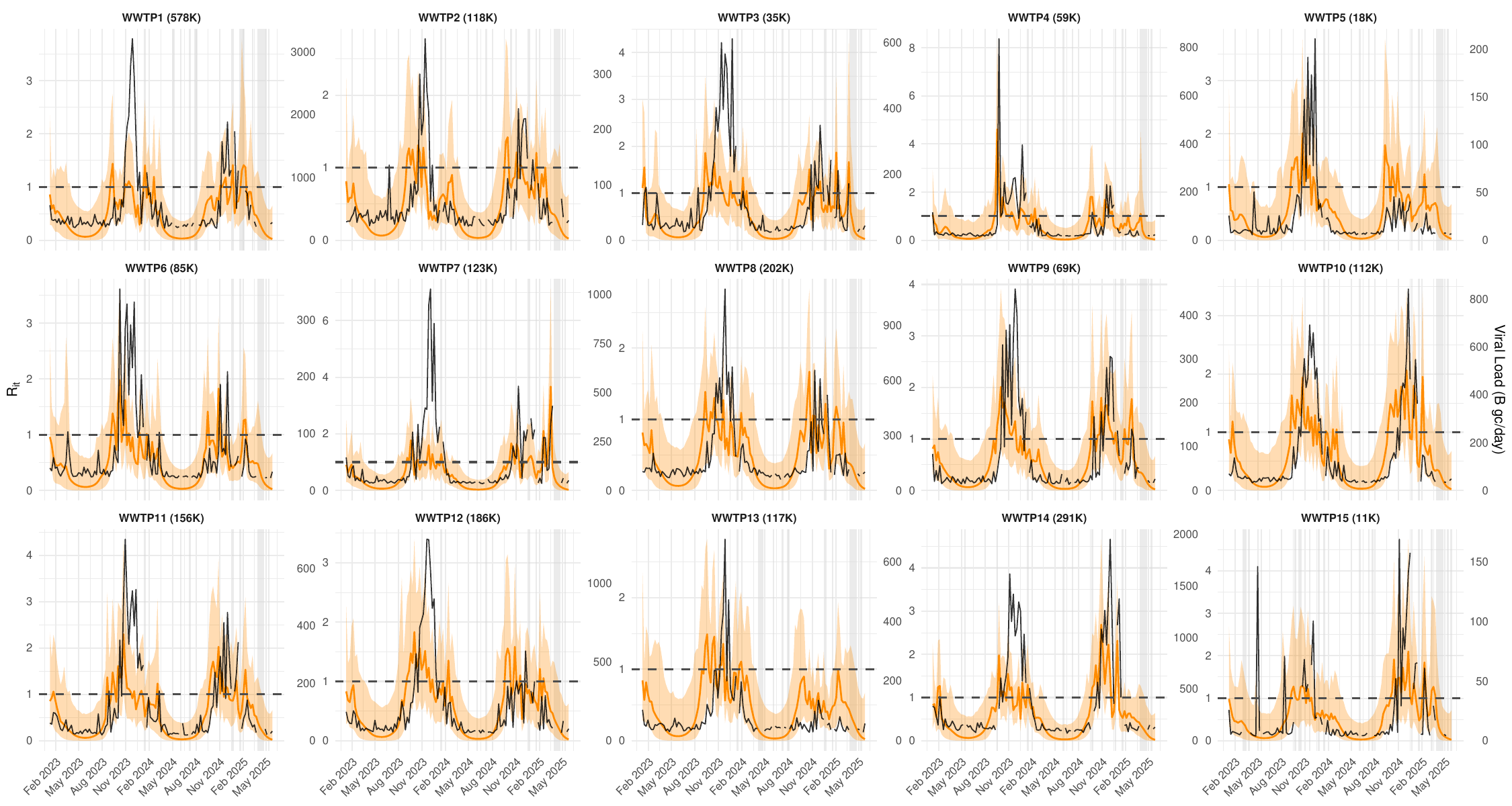}
\caption{Posterior $R_{it}$ under the second order extension of
the Filter variant with $\hat\rho$. Same panel layout and conventions as Figure \ref{fig:Rt-ssm-est}.}
\label{fig:Rt-ssm-ar2}
\end{figure}

\begin{figure}[H]
\centering
\includegraphics[width=\linewidth]{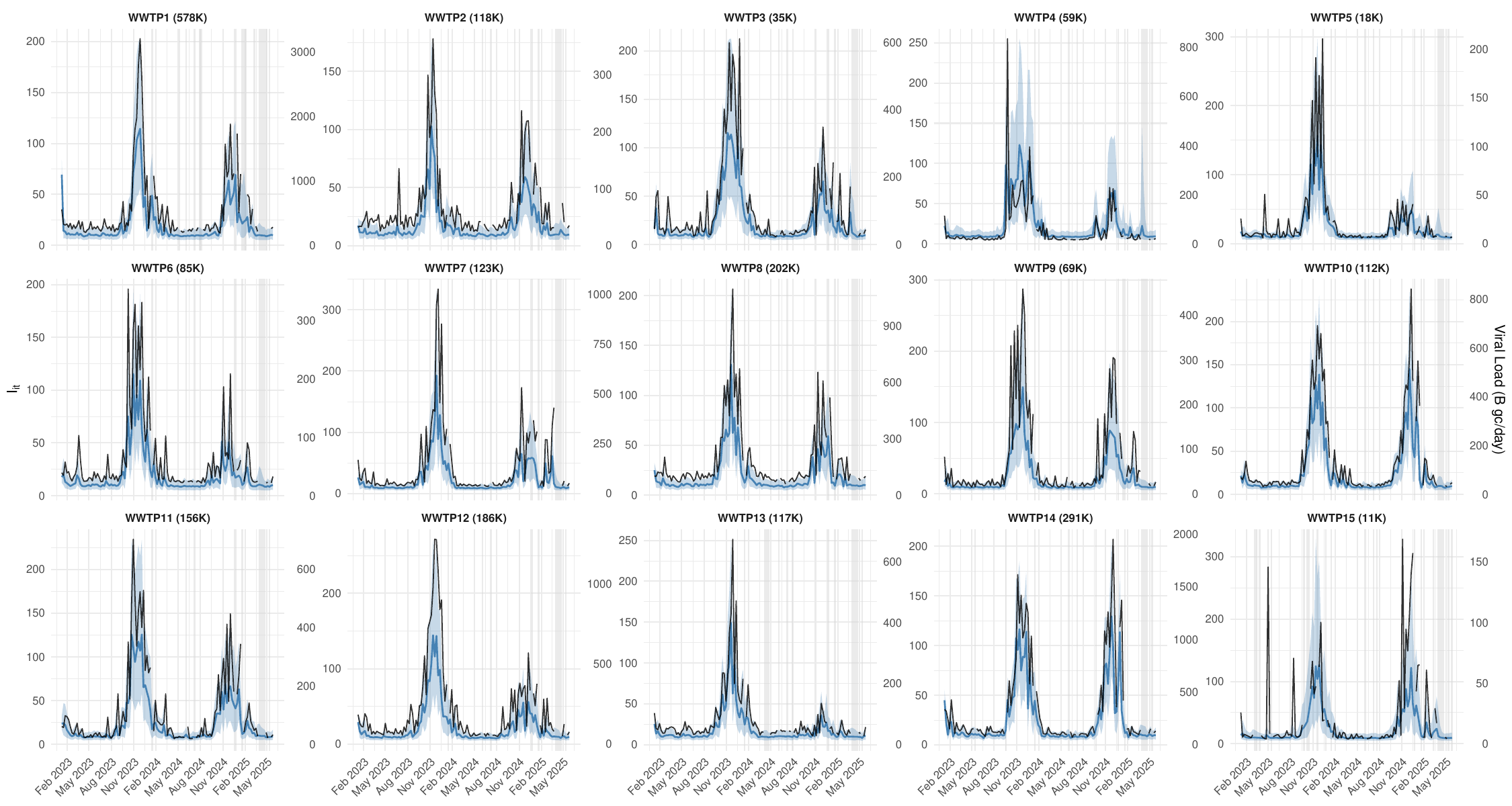}
\caption{Posterior $I_{it}$ under the second order extension of
the Filter variant with $\hat\rho$. Same panel layout and conventions as Figure \ref{fig:It-ssm-est}.}
\label{fig:It-ssm-ar2}
\end{figure}

\begin{figure}[H]
\centering
\includegraphics[width=\linewidth]{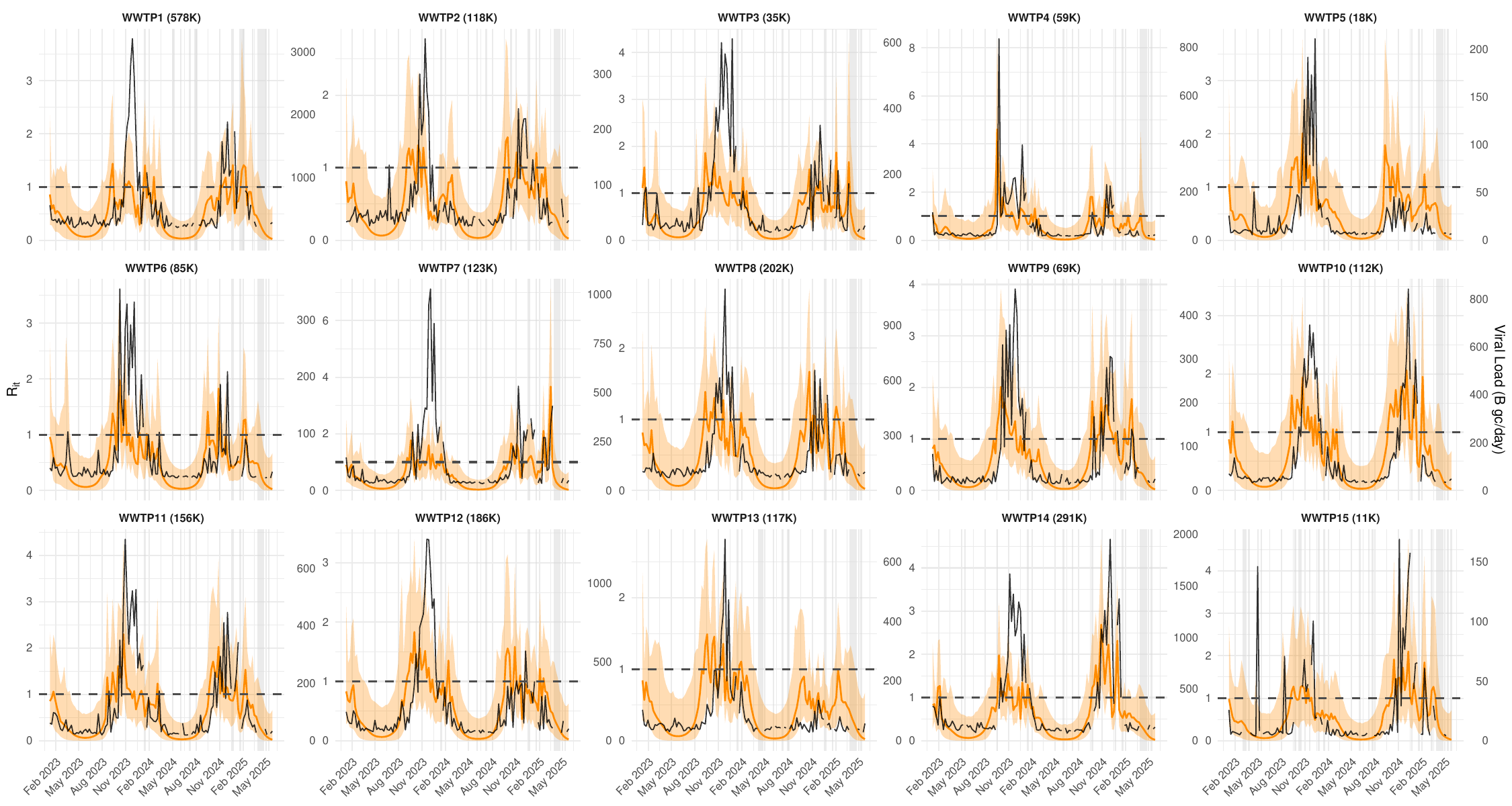}
\caption{Posterior $R_{it}$ under the second order extension of
the Data variant with $\hat\rho$. Same panel layout and conventions as Figure \ref{fig:Rt-direct-est}.}
\label{fig:Rt-direct-ar2}
\end{figure}

\begin{figure}[H]
\centering
\includegraphics[width=\linewidth]{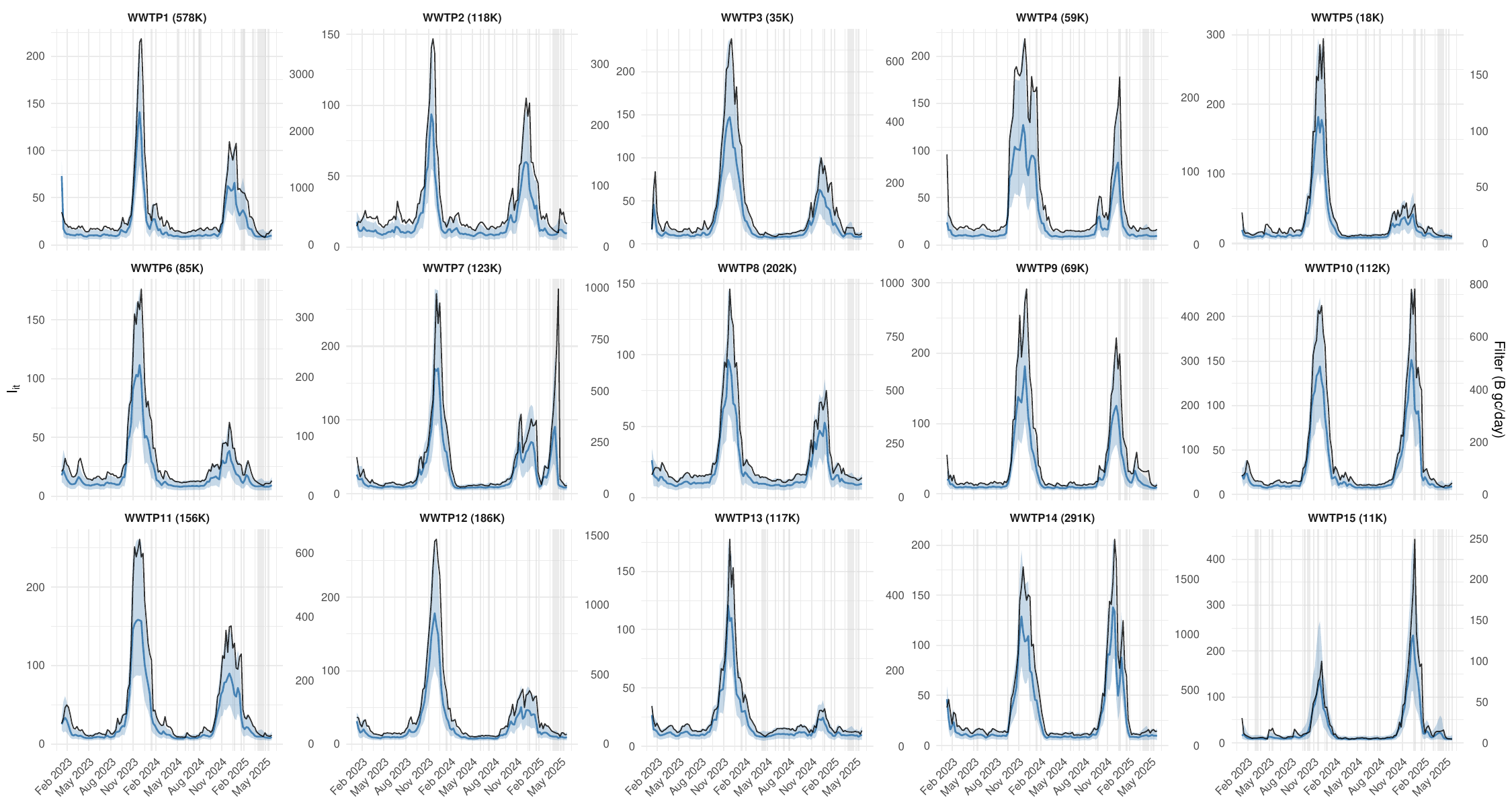}
\caption{Posterior $I_{it}$ under the second order extension of
the Data variant with $\hat\rho$. Same panel layout and conventions as Figure \ref{fig:It-direct-est}.}
\label{fig:It-direct-ar2}
\end{figure}

Figures \ref{fig:Rt-ssm-ar2}--\ref{fig:It-direct-ar2} report the posterior trajectories of $R_{it}$ and $I_{it}$ under the second-order extension for both the Filter and Data variants with $\hat\rho$. Qualitatively, the trajectories resemble those of the corresponding first-order variants (Figures \ref{fig:Rt-ssm-est} and \ref{fig:It-ssm-est} for Filter; Figures \ref{fig:Rt-direct-est} and \ref{fig:It-direct-est} for Data), but with smoother local behavior and a posterior of $\rho$ shifted upward ($\hat\rho \approx 0.67$ for Filter, $0.86$ for Data, versus $0.54$ and $0.76$ respectively in the first-order specification), reflecting the stronger persistence the second-order dynamics impose on the local trend.

\end{document}